\documentclass[prb,twocolumn,showpacs,preprintnumbers,amsmath,amssymb]{revtex4}

\usepackage{graphicx}
\usepackage{amsmath}
\usepackage{epstopdf}
\usepackage{amsbsy}
\usepackage{color}
\usepackage{dcolumn}
\usepackage{bm}
\usepackage[retainorgcmds]{IEEEtrantools}

\begin{document}

\title{Two routes to magnetic order by disorder in underdoped cuprates}
\author{R. B. Christensen$^1$, P. J. Hirschfeld$^2$, and B. M. Andersen$^1$}
\affiliation{$^1$Niels Bohr Institute, University of Copenhagen, Universitetsparken 5, DK-2100 Copenhagen,
Denmark\\
$^2$Department of Physics, University of Florida, Gainesville,
Florida 32611-8440, USA}

\date{\today}

\begin{abstract}

We study disorder-induced magnetism within the Gutzwiller approximation applied to the $t-J$ model relevant for cuprate superconductors.
In particular, we show how disorder generates magnetic phases by inducing local droplets of antiferromagnetic order which
eventually merge, and form a quasi-long range ordered state in the underdoped regime. We identify two distinct  disorder-induced magnetic phases of this type depending on the strength of the scatterers. For weak potential scatterers used to model dopant disorder, charge reorganization may push local regions in-between the impurities across the magnetic phase boundary, whereas for strong scatterers used to model substitutional ions, a local static magnetic moment is formed around each impurity. We calculate the density of states and find a remarkably universal low-energy behavior largely independent of both disorder and magnetization. However, the magnetic regions are characterized by larger (reduced) superconducting gap (coherence peaks) and a sub-gap kink in the density of states.

\end{abstract}


\maketitle

\section{Introduction}

Local experimental probes have highlighted the importance of intrinsic disorder and spatial inhomogeneity in the studies of high-Tc superconductors. In particular, scanning tunneling spectroscopy (STS) measurements have revealed nano-scale modulations in the gap for quasiparticle excitations and the local density of states (LDOS).\cite{cren,howald,lang,mcelroy} Complementary to these (energy resolved) density modulations, it is well-known that the spin density is also heterogeneous and exhibits spin-glass behavior in the underdoped regime. This seems to apply to both "clean" cuprates like YBa$_2$Cu$_3$O$_{6+x}$ (YBCO) where quasi-static SDW order is found in the far-underdoped regime,\cite{stock06, sonier,stock08,hinkov} and to intrinsically disordered materials like La$_{2-x}$Sr$_x$CuO$_4$ (LSCO) where the static spin correlations are long-range, and persist for a large doping range well into the superconducting dome.\cite{keimer,wakimoto,bella,julien} The size of the spin-glass phase in temperature and doping is clearly enhanced by disorder.  At present a detailed understanding is lacking of the connection between the modulated spin density and the STS tunneling data.\cite{Andersen08}

In order to further probe the effects of disorder, samples with additional extrinsic impurities have also been studied. Most notably, substitutional Zn ions strongly suppress T$_c$ and induce localized low-energy density of states resonances.\cite{pan} In the spin sector, Zn is known to slow-down and eventually freeze the spin fluctuations.\cite{hirota,bella,kimura03,savici} For example, in near-optimally doped LSCO increasing amounts of Zn substitution has been shown to shift spectral weight into the spin gap, and eventually generate elastic magnetic peaks in the neutron response.\cite{kimura03} A similar Zn-induced spin-freezing has been obtained for YBCO\cite{sidis96,suchaneck} and is generally consistent with $\mu$SR data on underdoped cuprates.\cite{mendels,bernhard,niedermayer,watanabe,panagopoulos}

Theoretical modeling of these experimental results are complicated by the need to include both charges, spins, and realistic disorder configurations.\cite{kaul,robertson,delmaestro,BMAndersen,BMAndersen:2006,alvarez,atkinson} Within an unrestricted Hartree-Fock approximation of the Hubbard model including $d$-wave pairing one may acquire a qualitative understanding of the disorder-induced magnetism; due to the splitting of low-energy in-gap impurity states, it can be advantageous to generate local AF order which may percolate, and eventually form a quasi-long-range ordered state.\cite{BMAndersen,Schmid10,AndersenJPCS} This scenario is a many-impurity generalization of the well-known one-impurity result of induced local magnetization,\cite{tsuchiura,wanglee,zhu,chen04,BMAndersen06,JWHarter:2007,HAlloul:2009} and agrees with transport measurements.\cite{andersen08,weichen} Recently, the dynamics of spin freezing by impurities, i.e. the shift of magnetic spectral weight to low energies, were calculated explicitly and shown to be consistent with this scenario as well.\cite{andersen10}

The above approach, while successful to an extent, has obvious theoretical drawbacks.  First, it cannot describe the approach to the Mott insulator; underdoping has to be understood as the effective increase of correlations represented by $U/t$ as one underdopes due to the suppression of screening, but there is no way to calculate this effect systematically.  Secondly, the connection between correlations and pairing is entirely artificial, since the BCS pairing term is added by hand and treated in mean field.  What is needed is a technique which allows for the study of local variations of observable quantities in the presence of disorder, which easily accounts at least for the crude effects of strong correlations in the underdoped regime.

Here, we study disorder-induced magnetism within the Gutzwiller approximation (GA) of the $t-J$ model. The GA has not been traditionally applied to include spin non-degeneracy, but was extended to include antiferromagnetism by Ogata and Himeda for the homogeneous case.\cite{Himeda1,Ogata1} The so-called extended Gutzwiller factors (EGFs) have been applied to inhomogeneous problems by interpreting them in a site-dependent manner.  The site-dependent EGFs have been used to study local magnetic moments around a nonmagnetic impurity,\cite{tsuchiura} electronic states around a vortex core,\cite{tsuchiura2} and anti-phase superconducting domain structures.\cite{tsuchiura3} More recently, a simplified version of the EGFs, with the advantage that they reduce to the well-defined extensively tested original site-dependent Gutzwiller factors,\cite{Wing-Ho} have been used to examine the energetics of the charge and spin stripe ordered superconducting state.\cite{Yang} We focus on two distinct cases: 1) weak scatterers with impurity concentration equal to the doping level, $n_{imp} = \delta$, modeling the disorder potential from out-of-plane dopants, and 2) small concentrations, $n_{imp}=1\!-\!2\%$, of strong scatterers simulating the effects of substitutional disorder e.g. Zn or vacancies. In both cases, we find that disorder may induce AF phases in the underdoped regime. The origin of the induced magnetism is vastly different, however. In the former case of dopant disorder, the redistributed charge density creates local regions close to half-filling in-between the impurities, pushing these areas across the bulk magnetic phase boundary. We denote this kind of magnetic phases by type I. By contrast, in the other case of substitutional disorder, even a single impurity nucleates magnetization in its vicinity, and the final spin structure consists of overlapping single-impurity regions which may orient themselves in order to minimize the exchange energy.\cite{BMAndersen} This kind of disorder-induced magnetic phase is called type II in this paper. The density of states in the disordered phase largely retains its $d$-wave 'V' shape in agreement with experiments\cite{mcelroy,fang,gomes} and consistent with earlier theoretical studies.\cite{BMAndersen:2006,andersen08,garg,tklee2009} Surprisingly, however, the magnetic regions exhibit a larger superconducting gap and reduced associated coherence peaks in agreement with the general characteristics of the large-gap regions in the experimental STS data. Finally we discuss additional sub-gap features in the LDOS which could function as fingerprints for local magnetism in tunneling experiments.

\section{Model}
The $t-J$ Hamiltonian, defined on a 2D lattice, is given by
\begin{equation}
H_{t-J}=-\sum_{(ij)\sigma}P_{G}(t_{ij}\hat{c}_{i\sigma}^{\dagger}\hat{c}_{j\sigma}+\mbox{H.c.})P_{G}+J\sum_{\langle ij \rangle}\hat{\textbf{S}}_{i}\cdot \hat{\textbf{S}}_{j}, \label{t-J}
\end{equation}
where $c_{i\sigma}^\dagger$ creates an electron at site $i$ with spin $\sigma$. $\textbf{S}_{i}$ is the spin operator for site $i$ and $P_{G}$ is the Gutzwiller projector defined by $P_{G}=\prod_{i}(1-\hat{n}_{i\uparrow}\hat{n}_{i\downarrow})$, where $\hat{n}_{i\sigma}=\hat{c}_{i\sigma}^{\dagger}\hat{c}_{i\sigma}$ is the spin dependent number operator. For all results shown in this paper we have used $J/t=0.3$ and $t'=-0.25t$. In Eq. \eqref{t-J}, $\langle ij \rangle$ denotes nearest-neighbor sites with associated hopping amplitude $t$ whereas $(ij)$ refers to both nearest- and next-nearest neighbor sites with hopping amplitudes $t$ and $t'$, respectively. Disorder is introduced into the system by $N$ point-like scatterers
\begin{eqnarray}
H_{imp}=\sum_{i}V_{i}\hat{n}_{i}.
\end{eqnarray}
To solve the $t-J$ model, the no double occupancy constraint has to be approximated. Zhang {\it el al.}\cite{Zhang1} introduced the Gutzwiller approximation (GA) to replace the Gutzwiller projectors in Eq. \eqref{t-J}; in this paper we use a  simplified version of the EGFs giving  rise to the following renormalized Hamiltonian
\begin{align}
& H = -\sum_{(ij)\sigma}g^{t}_{ij}t_{ij}\left(\hat{c}_{i\sigma}^{\dagger}\hat{c}_{j\sigma}+\mbox{H.c}\right) \nonumber \\
& + \sum_{\langle ij\rangle } J\left[g^{s,z}_{ij}\hat{S}^{s,z}_{i}\hat{S}^{s,z}_{j}+g_{ij}^{s,xy}\left(\frac{\hat{S}^{+}_{i}\hat{S}^{-}_{j}+\hat{S}^{-}_{i}\hat{S}^{+}_{j}}{2} \right)\right] \nonumber \\
& +\sum_{i}V_{i}\hat{n}_{i}. \label{TJH}
\end{align}
The simplified EGFs depend on the local values of the magnetic and pairing order parameters, the local kinetic energy, and hole density defined  by
\begin{IEEEeqnarray}{rCl}\label{mdef}
m_{i}&=&\langle\Psi_{0}| \hat{S}_{i}^{z}|\Psi_{0}\rangle,  \\\label{deltadef}
\Delta_{ij\sigma}&=&\sigma\langle\Psi_{0}| \hat{c}_{i\sigma}\hat{c}_{j\bar{\sigma}}|\Psi_{0}\rangle,  \\\label{chidef}
\chi_{ij\sigma}&=&\langle\Psi_{0}| \hat{c}_{i\sigma}^{\dagger}\hat{c}_{j\sigma}|\Psi_{0}\rangle,  \\
\delta_{i}&=&1-\langle\Psi_{0}|\hat{n}_{i}|\Psi_{0}\rangle,  \label{molecular fields}
\end{IEEEeqnarray}
where $|\Psi_{0}\rangle$ denotes the unprojected ground state wave function.
The simplified EGFs are given as
\begin{IEEEeqnarray}{rCl}\label{SEGF1}
g^{t}_{ij\sigma}&=& g^{t}_{i\sigma}g^{t}_{j\sigma},  \\
g^{t}_{i\sigma}&=& \sqrt{\frac{2\delta_{i}(1-\delta_{i})}{1-\delta_{i}^{2}+4m^{2}}\frac{1+\delta_{i}+\sigma2m_{i}}{1+\delta_{i}-\sigma2m_{i}}},  \\
g^{s,xy}_{ij}&=& g^{s,xy}_{i}g^{s,xy}_{j},  \\
g^{s,xy}_{i}&=& \frac{2(1-\delta_{i})}{1-\delta_{i}^{2}+4m_{i}^{2}},  \\
g^{s,z}_{ij} &=& g^{s,xy}_{ij}\frac{2(\bar{\Delta}_{ij}^{2}+\bar{\chi}_{ij}^{2})-4m_{i}m_{j}X^{2}_{ij}}{2(\bar{\Delta}_{ij}^{2}+\bar{\chi}_{ij}^{2})-4m_{i}m_{j}},  \\
X_{ij}&=&1+\frac{12(1-\delta_{i})(1-\delta_{j})(\bar{\Delta}_{ij}^{2}+\bar{\chi}_{ij}^{2})}{\sqrt{(1-\delta_{i}^{2}+4m_{i}^{2})(1-\delta_{j}^{2}+4m_{j}^{2})}}, \label{SEGF}
\end{IEEEeqnarray}
where $\bar{\Delta}_{ij}=\sum_{\sigma} \frac{\Delta_{ij\sigma}}{2}$ and $\bar{\chi}_{ij}=\sum_{\sigma} \frac{\chi_{ij\sigma}}{2}$.
Note that the simplified EGFs allow for $\Delta_{\uparrow}\neq\Delta_{\downarrow}$.
The rewriting in Eqs.\eqref{SEGF1}-\eqref{SEGF} of the EGFs is identical to that used by Yang {\it et al.}\cite{Yang}

A direct diagonalization of the Hartree-Fock Hamiltonian $H_{H-F}$ obtained from a mean-field decoupling in Eq. \eqref{TJH} is not sufficient because the simplified EGFs also depend on the order parameters. Instead, the energy has to be calculated from the mean field Hartree-Fock Hamiltonian and then minimized with respect to the unprojected wave function $|\Psi_{0}\rangle$ under the constraints of both fixed total electron density $\sum_{i}n_{i}=N_{e}$, and fixed wavefunction normalization $\langle\Psi_{0}|\Psi_{0}\rangle=1$.\cite{Yang} This is equivalent to minimizing the function:
\begin{IEEEeqnarray}{rCl}\nonumber
W&=&\langle\Psi_{0}|H_{H-F}|\Psi_{0}\rangle -\lambda(\langle\Psi_{0}|\Psi_{0}\rangle-1)
\\
&&-\mu\left(\sum_{i}\hat{n}-N_{e}\right),
\end{IEEEeqnarray}
which leads to the following renormalized mean-field Hamiltonian
\begin{align}
&H_{\mbox{mf}}=\sum_{(ij)\sigma} \frac{\partial W}{\partial\chi_{ij\sigma}}\hat{c}_{i\sigma}^{\dagger}\hat{c}_{j\sigma}+\mbox{H.c.}\\
&+\sum_{\langle ij\rangle \sigma} \frac{\partial W}{\partial\Delta_{ij\sigma}}\sigma\hat{c}_{i\sigma}\hat{c}_{j\bar{\sigma}}+\mbox{H.c.}+\sum_{i\sigma} \frac{\partial W}{\partial\hat{n}_{i\sigma}}\hat{n}_{i\sigma}, \nonumber
 \label{H-final}
\end{align}
\begin{widetext}
with the self-consistent equations
\begin{IEEEeqnarray}{rCl}
\frac{\partial W}{\partial\chi_{ij\sigma}}&=&-\delta_{ij,\langle ij \rangle }J\left(\frac{g_{ij }^{s,z}}{4}+\frac{g_{ij }^{s,xy}}{2}\frac{\chi_{ij \bar{\sigma}}^{*}}{\chi_{ij \sigma}^{*}}\right)\chi_{ij \sigma}^{*}-g_{ij\sigma}t_{ij}
\\
&&-\frac{J}{4}\left(|\Delta_{ij\uparrow}|^{2}+|\Delta_{ij\downarrow}|^{2}+|\chi_{ij\uparrow}|^{2}+|\chi_{ij\downarrow}|^{2}-4m_{i}m_{j}\right)\frac{dg_{ij }^{s,z}}{d\chi_{ij \sigma}},  \nonumber\\
\frac{\partial W}{\partial\Delta_{ij \sigma}}&=&-J\left(\frac{g_{ij }^{s,z}}{4}+\frac{g_{ij }^{s,xy}}{2}\frac{\Delta_{ij \bar{\sigma}}^{*}}{\Delta_{ij \sigma}^{*}}\right)\Delta_{ij \sigma}^{*} \label{dWdDelta} \\
&&-\frac{J}{4}\left(|\Delta_{ij\uparrow}|^{2}+|\Delta_{ij\downarrow}|^{2}+|\chi_{ij\uparrow}|^{2}+|\chi_{ij\downarrow}|^{2}-4m_{i}m_{j}\right)\frac{dg_{ij }^{s,z}}{d\Delta_{ij \sigma}}, \nonumber \\
\frac{\partial W}{\partial n_{i\sigma}}&=&-\left(\mu -V_{i}\right) +\frac{1}{2}\sigma \sum_{j}g_{ij }^{s,z}J m_{j} \\  &&-\frac{J }{4}\sum_{j}\left(|\Delta_{ij\uparrow}|^{2}+|\Delta_{ij\downarrow}|^{2}+|\chi_{ij\uparrow}|^{2}+|\chi_{ij\downarrow}|^{2}-4m_{i}m_{j}\right)\frac{dg_{ij }^{s,z}}{dn_{i\sigma}}, \nonumber
\\
&&-\frac{J}{2}\sum_{j\sigma'}(\left(\chi_{ij \bar{\sigma}'}^{*}\chi_{ij \sigma'} +\Delta_{ij \bar{\sigma}'}^{*}\Delta_{ij \sigma'}\right))\frac{dg_{ij}^{s,xy}}{dn_{i\sigma}}- \sum_{j\sigma'} t_{ij}\frac{dg_{ij\sigma'}^{t}}{dn_{i\sigma}}\left(\chi_{ij\sigma'}+\chi_{ij\sigma'}^{*} \right). \nonumber \label{SelfConsisten}
\end{IEEEeqnarray}
\end{widetext}
Here $\bar{\sigma}$ denotes the opposite spin of $\sigma$.
The derivatives of the EGFs entering these equations can be straightforwardly derived from Eqs.\eqref{SEGF1}-\eqref{SEGF}. We have solved these unrestricted equations self-consistently by iteration on $24\times 24$ lattices, by diagonalization of the Bogoliubov-de Gennes (BdG) equations associated with the excitation operators $\hat{\gamma}_{n\sigma}^{\dagger}$ and $\hat{\gamma}_{n\sigma}$ defined by $\hat{c}_{i\uparrow} = \sum_{n} \left( u_{ni\uparrow}\hat{\gamma}_{n\uparrow}+v^{*}_{ni\uparrow}\hat{\gamma}^{\dagger}_{n\downarrow} \right)$.\cite{JWHarter:2007,Barash}

\begin{figure}[h!]
\includegraphics[clip=true,width=0.98\columnwidth]{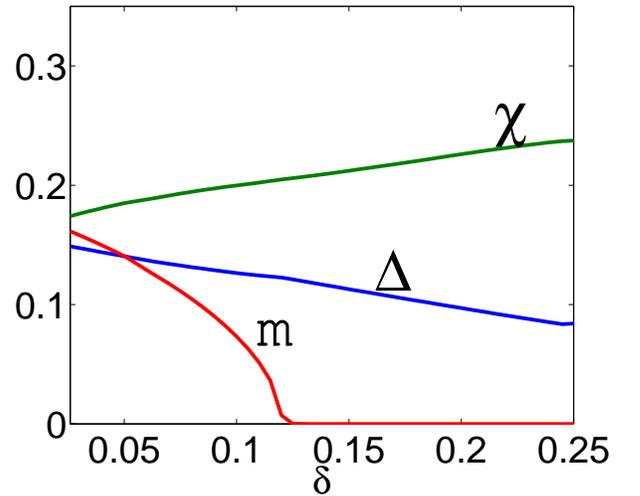}
\caption{(Color online) Phase diagram for the homogeneous case with commensurate $(\pi,\pi)$ AF order $m$, $d$-wave superconductivity $\Delta$, and kinetic energy $\chi$ [see Eqs.(\ref{mdef}-\ref{chidef})].} \label{phasediagram}
\end{figure}

At low doping, such a simple iterative procedure fails to converge in the inhomogeneous case because $\frac{dg_{i}^{t}}{dn_{i\sigma}}$ diverges when the local density approaches half-filling, $\delta = 0$. Therefore, small changes in the electron density between consecutive iterations causes large jumps in $\frac{dg_{i}^{t}}{dn_{i\sigma}}$. In this paper, we therefore restrict the doping level of the inhomogeneous cases to be larger that $\delta>0.115$ where iterations are sufficient for obtaining numerical convergence. On the other hand, for the homogeneous case allowing for a superconducting and a commensurate AF phase, the phase diagram can be easily mapped out for $\delta>0.026$  and is shown in Fig. \ref{phasediagram}. This phase diagram is similar to that obtained e.g. in Refs. \onlinecite{Ogata1,Chen}. Note that the phase boundary for the magnetic order in Fig. \ref{phasediagram} is valid only for a standard $(\pi,\pi)$ AF order. At finite doping striped magnetic order can also be stabilized within a similar approach,\cite{Yang} exhibiting a slightly different phase boundary.

\section{Results}

We begin the results section by discussing the case of dopant disorder, i.e. $n_{imp} = \delta$ where each scatterer is relatively weak, $V_i=t$. In previous studies of unrestricted Hartree-Fock applied to the Hubbard model it was found that each dopant induced local magnetization, leading to a scenario where the amount of disorder-induced magnetization is proportional to the doping level contrary to experiments.\cite{BMAndersen} Such an approach, however, does not include any band-widening with increased doping, and can be made consistent with the lack of magnetization in the overdoped regime only by requiring $U/t$ to be a decreasing function of doping. The present model naturally includes the effects of strong correlations in the underdoped regime and, as seen from Fig. \ref{figmagn}, the dopant disorder indeed induces a finite magnetization but only at low doping levels. The magnetic phase shown in Fig. \ref{figmagn}, which we denote type I, is incommensurate as seen from Fig. \ref{figx0125}(a), and disorder-induced as verified by a vanishing magnetization in the absence of disorder (not shown). The origin of the magnetization is a charge redistribution caused by the weak impurities: it is energetically favorable for the electrons to be located away from the disorder sites which then push these local regions across the magnetic phase boundary similar to the homogeneous case shown in Fig. \ref{phasediagram}. Clearly, this mechanism is dominant at low doping where regions more readily reach the local critical doping level.

\begin{widetext}

\begin{figure}[]
\includegraphics[clip=true,width=0.24\columnwidth]{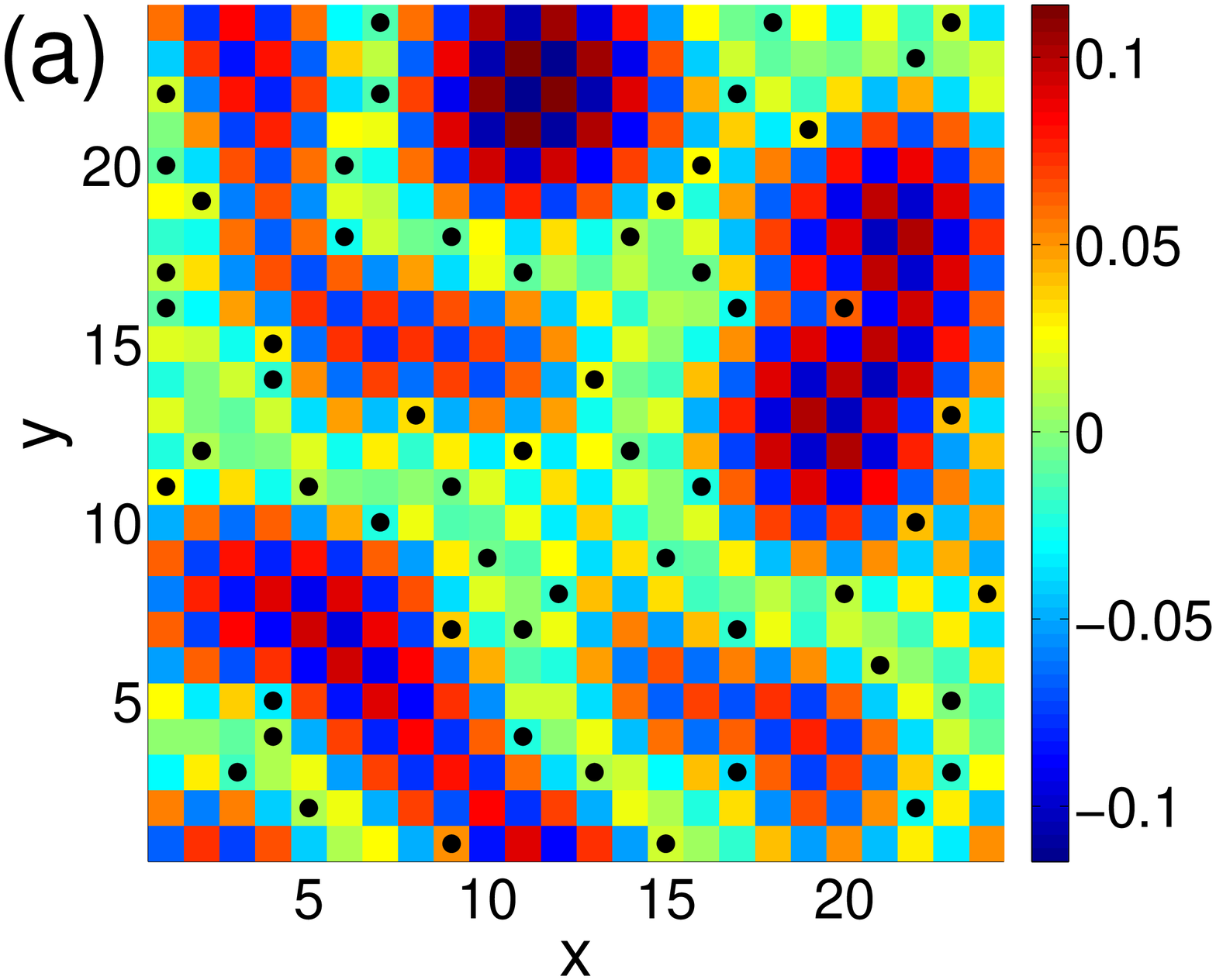}
\includegraphics[clip=true,width=0.24\columnwidth]{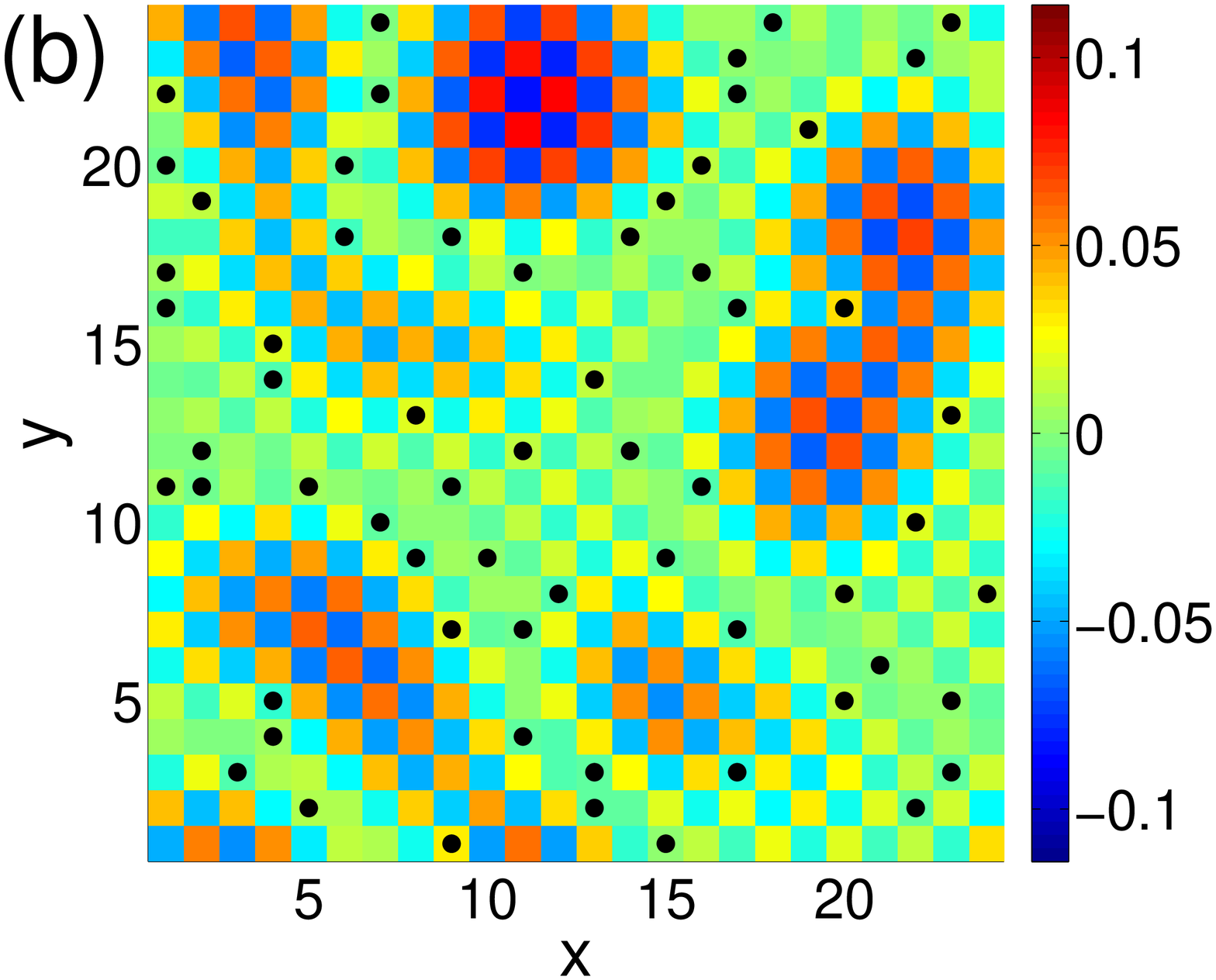}
\includegraphics[clip=true,width=0.24\columnwidth]{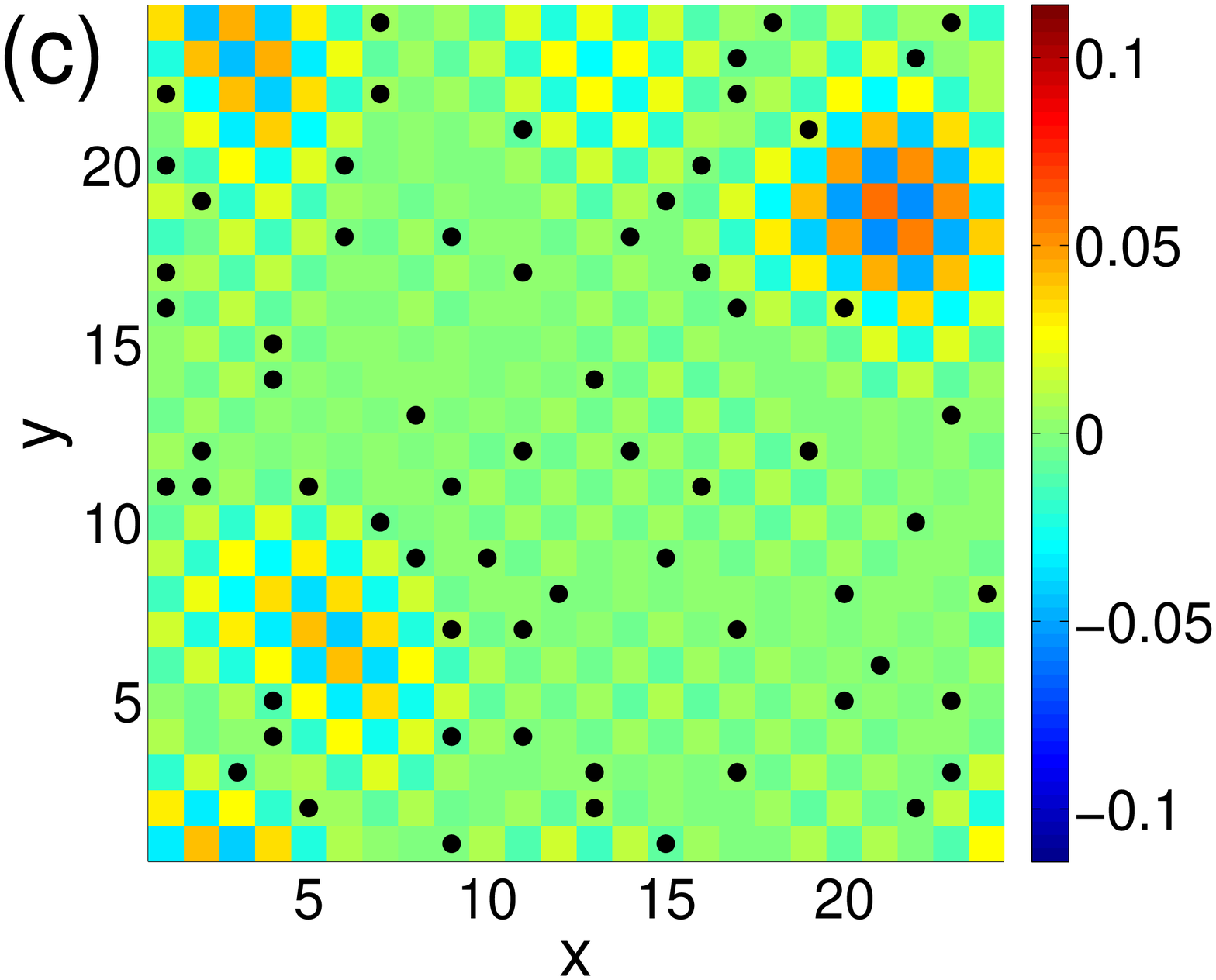}
\includegraphics[clip=true,width=0.24\columnwidth]{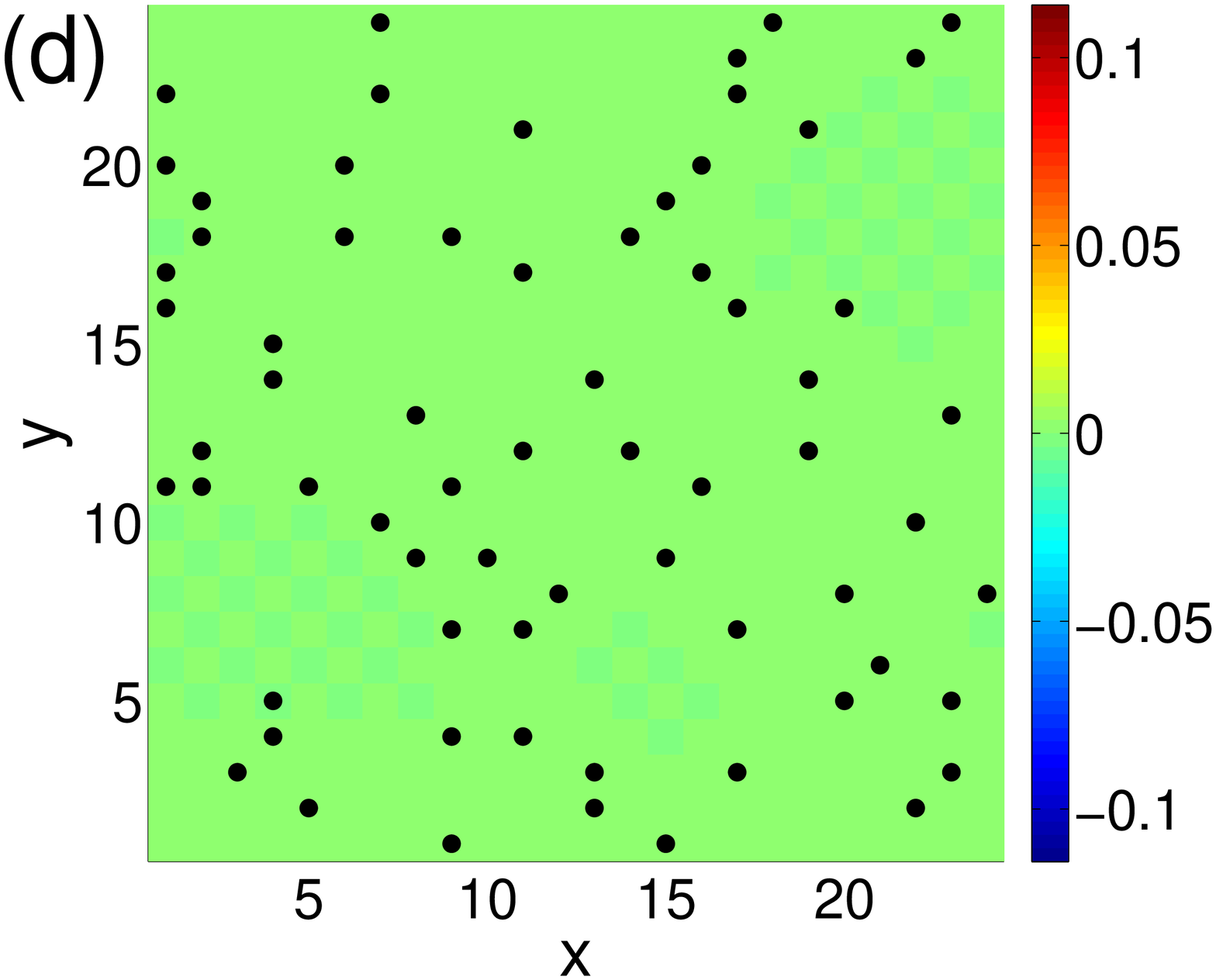}
\caption{(Color online) Type I behavior. Magnetization shown in a real-space field of view for doping levels close to the magnetic phase transition of the clean system $\delta=0.115, 0.125, 0.13, 0.135$ (left to right). The black dots show the positions of the impurities. The dopant disorder is modeled with a potential strength $V_i=t$, and $n_{imp}=\delta$ where $n_{imp}$ is the impurity concentration and $\delta$ the doping level. As seen, the dopant-induced magnetic regions gradually disappear as the doping increases.}
\label{figmagn}
\end{figure}

\begin{figure}[]
\includegraphics[clip=true,width=0.24\columnwidth]{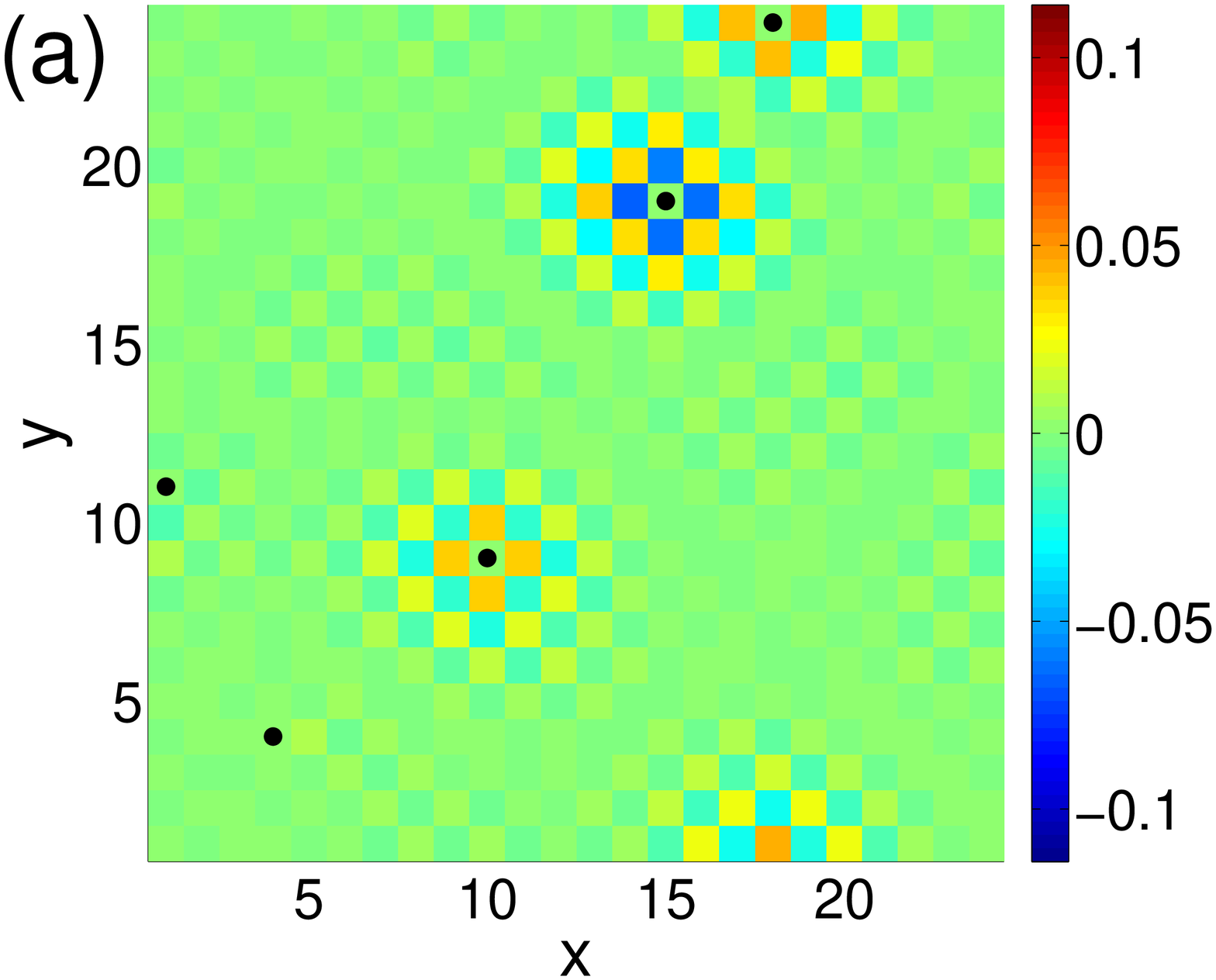}
\includegraphics[clip=true,width=0.24\columnwidth]{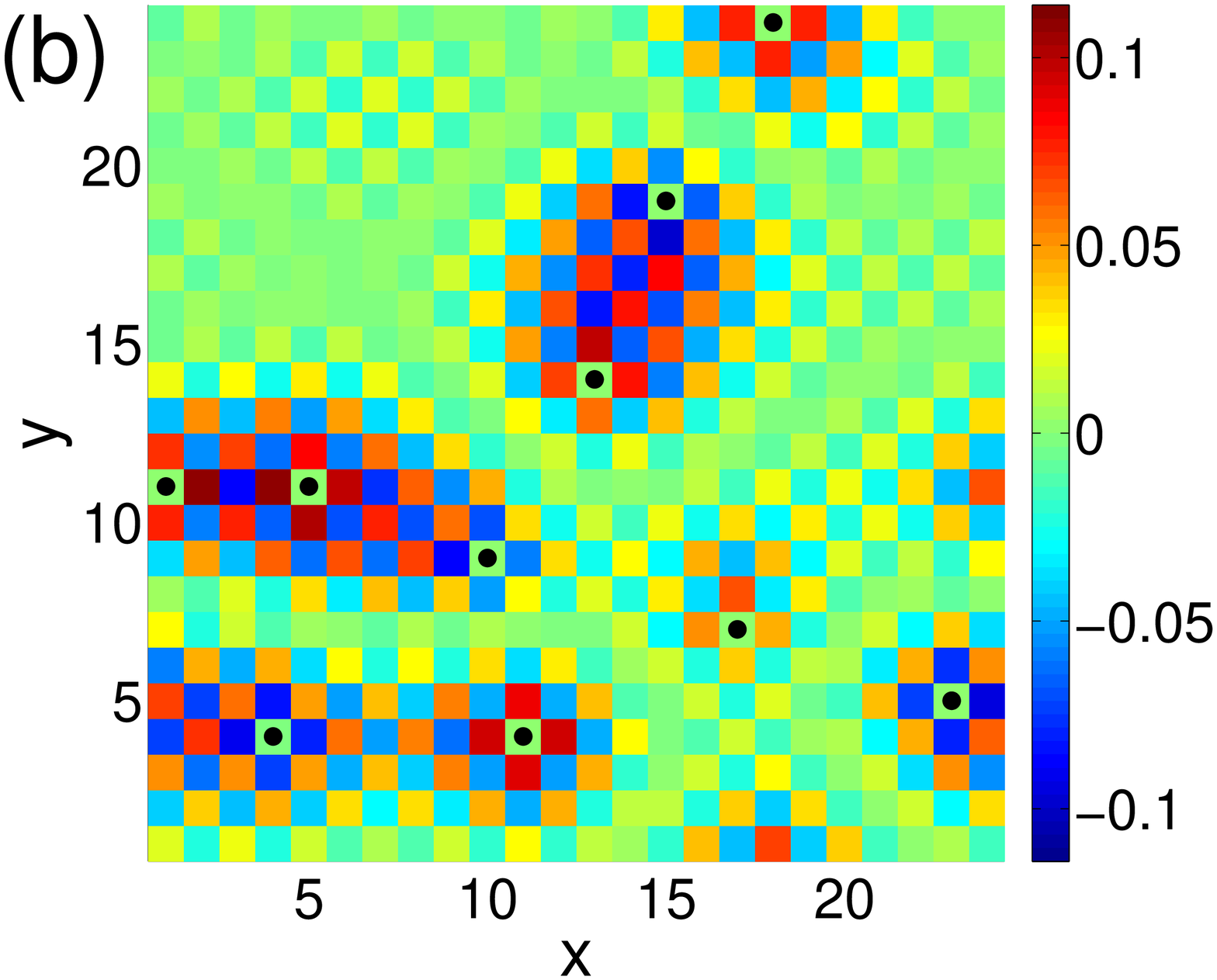}
\includegraphics[clip=true,width=0.24\columnwidth]{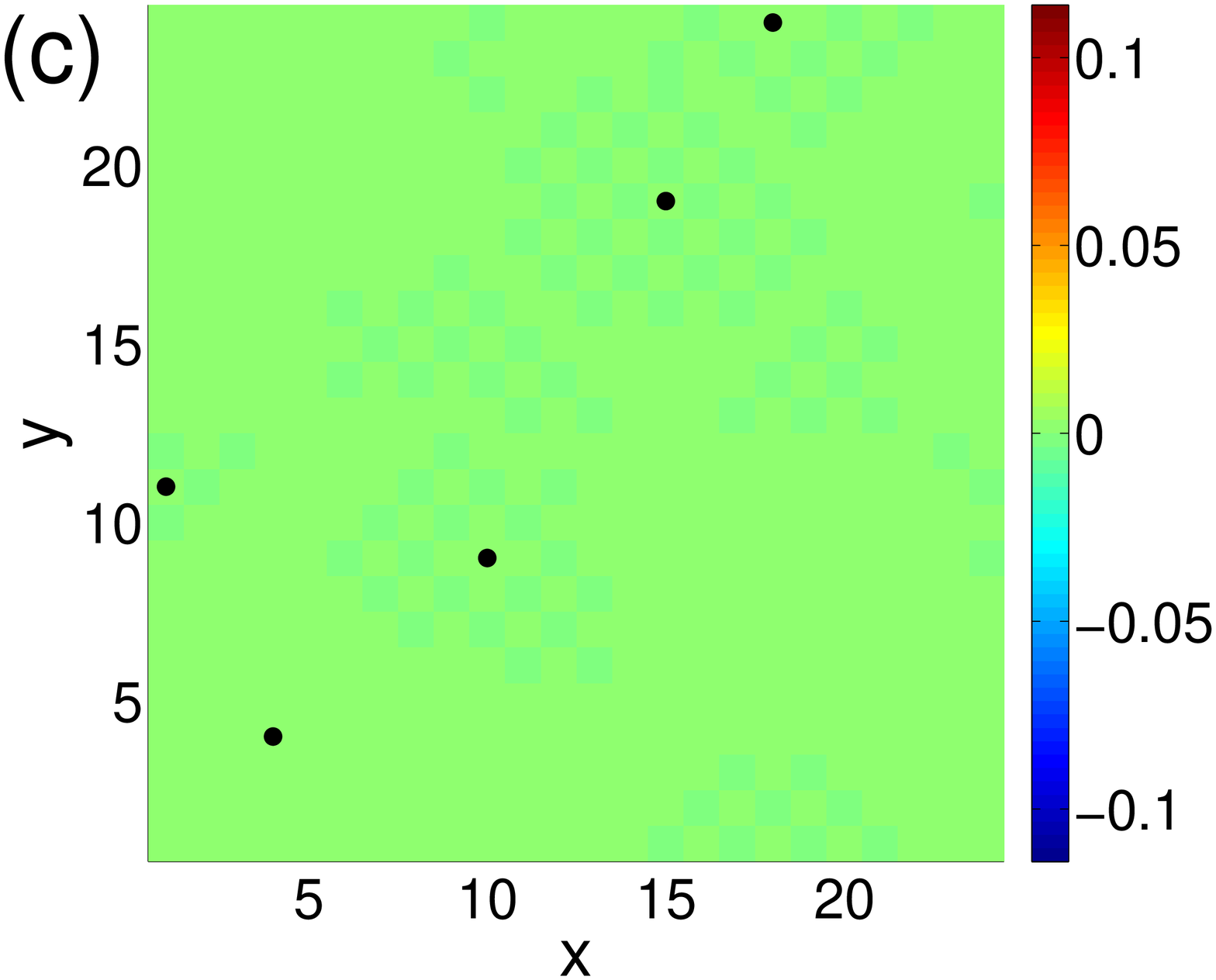}
\includegraphics[clip=true,width=0.24\columnwidth]{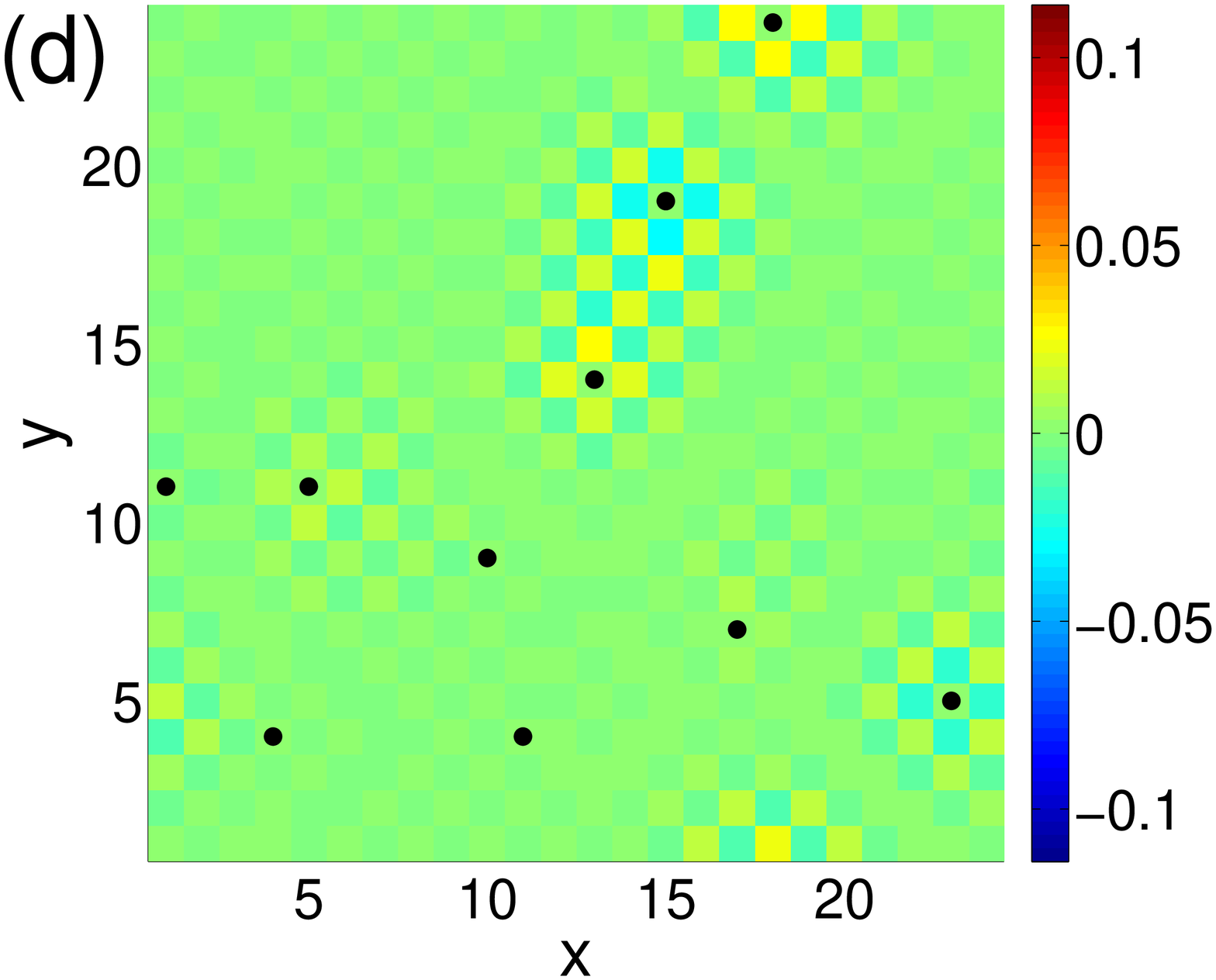}
\caption{(Color online) Type II behavior. Magnetization shown in a real-space field of view in the presence of $1\%$ (a,c) and $2\%$ (b,d) strong impurities with $V_i=100t$ for doping $\delta=0.14$ (a,b) and $\delta=0.15$ (c,d). }
\label{figmagnstrong}
\end{figure}

\end{widetext}

The origin of the magnetization in Fig. \ref{figmagn} is different from the disorder-induced AF studied previously, which we denote by type II, where each impurity gives rise to a local magnetization.\cite{tsuchiura,wanglee,zhu,chen04,BMAndersen06,BMAndersen,JWHarter:2007,HAlloul:2009} In the single impurity case, each defect gradually "freezes" magnetic fluctuations as correlations or the impurity potential increase, shifting spectral weight from high to low energies, eventually accumulating in a local $\omega=0$ peak. Within the present model, we reproduce the latter mechanism in the limit of strong scatterers as shown in Fig. \ref{figmagnstrong}. Specifically, Fig. \ref{figmagnstrong} shows a situation with $V_i=100t$ for $n_{imp}=1\%$(a,c) and $n_{imp}=2\%$(b,d), and doping $\delta=0.14$(a,b) and $\delta=0.15$ (c,d). Though care must be taken when extrapolating single-impurity to many-impurity effects in $d$-wave superconductors,\cite{chen04,Schmid10,atkinson2000,anderseninterference,morr} the results shown in Fig. \ref{figmagnstrong} can be explained from overlapping single-impurity magnetizations. The result agrees with the general notion that enough strong scatterers at a given doping level can lead to static magnetic order. For higher doping levels larger concentrations of impurities are needed to freeze the spins as shown explicitly by comparing e.g. Figs. \ref{figmagnstrong}(a) and \ref{figmagnstrong}(c).\cite{sidis96,kimura03,suchaneck,andersen10}

Next we focus on the result shown in Fig. \ref{figmagn}(a). Figure \ref{figx0125} shows other relevant physical quantities for this parameter set: (a) the Fourier transform of the magnetization, (b) the absolute value of the magnetization $|m_i|$, (c) the electronic charge density $n_i$, (d) the gapmap extracted from the LDOS, (e) the peak-height extracted from the LDOS, and (f) the gapmap extracted from the LDOS without the possibility for magnetic order. A comparison of Figs. \ref{figx0125}(b,c) verifies the direct correlation between the local density and the induced magnetization for type I behavior. Hence, the origin of the magnetization is local phase transitions caused by charge modulations with regions closer to half-filling pushed  across the magnetic phase boundary [see Fig. \ref{phasediagram}].

The LDOS $N_i(\omega)$ can be obtained from
\begin{equation}\label{Nomega}
N_i(\omega)=\sum_{n} g^{N}_{i\sigma} \left[ |u_{ni\sigma}|^{2} \delta(\omega-E_{n})+|v_{ni\sigma}|^{2} \delta(\omega+E_{n}) \right],
\end{equation}
where $i$ denotes the site index, $n$ is the index of the eigenstates with BdG eigenvalue $E_n$, and $g^{N}_{i\sigma}=\frac{\delta_i}{1-n_{i\sigma}}$ is the Gutzwiller renormalization factor originating from the "hopping" between different times entering Eq. \eqref{Nomega}.\cite{garg} Surprisingly, the local superconducting gap shown in Fig. \ref{figx0125}(d) extracted from the LDOS is {\it larger} in the magnetic regions of the system contrary to the expectation from a conventional competitive scenario. The larger gap results from a combined effect of 1) lower effective doping in the magnetic regions and hence a larger local pairing order parameter [see Fig. \ref{phasediagram}] caused by the impurity induced redistribution of the electron-density, which is enhanced by the magnetic order, and 2) a renormalization of the local pairing constant in the presence of magnetization dominated by the last term proportional to $m_im_j$ in Eq. \eqref{dWdDelta} (see also Fig. \ref{fig:J} below).  Comparing the gapmap with and without magnetic order in Fig. \ref{figx0125}(d) and Fig. \ref{figx0125}(f) clearly illustrates that magnetic order contributes significantly to the spatial modulation of the gap measured by STS in the underdoped regime.\cite{cren,howald,lang,mcelroy} However the two bulk arguments above are not  sufficient to explain the entire modulation in Fig. \ref{figx0125}(d)  indicating a nontrivial effect of the spatial inhomogeneity.

The peak height of the coherence peaks also varies in real space and tends to be anti-correlated with the local gap magnitude as seen from Fig. \ref{figx0125}(e). This anti-correlation is a direct consequence of $g^N_{i\sigma}$ in Eq. \eqref{Nomega} which reduces the low-energy spectral weight in large gap regions due to their closer proximity to half-filling. The anti-correlation between local gap and coherence peak height is consistent with STM measurements\cite{mcelroy2005} which motivated a picture of locally modulated pairing in the cuprate superconductors.\cite{nunner1,nunner2,andersentherm} Within the present strong-coupling approach, we find a similar anti-correlation for type I disorder-induced magnetization in the underdoped regime.

 \begin{figure}[]
\begin{minipage}{.49\columnwidth}
\includegraphics[clip=true,height=0.98\columnwidth,width=0.98\columnwidth]{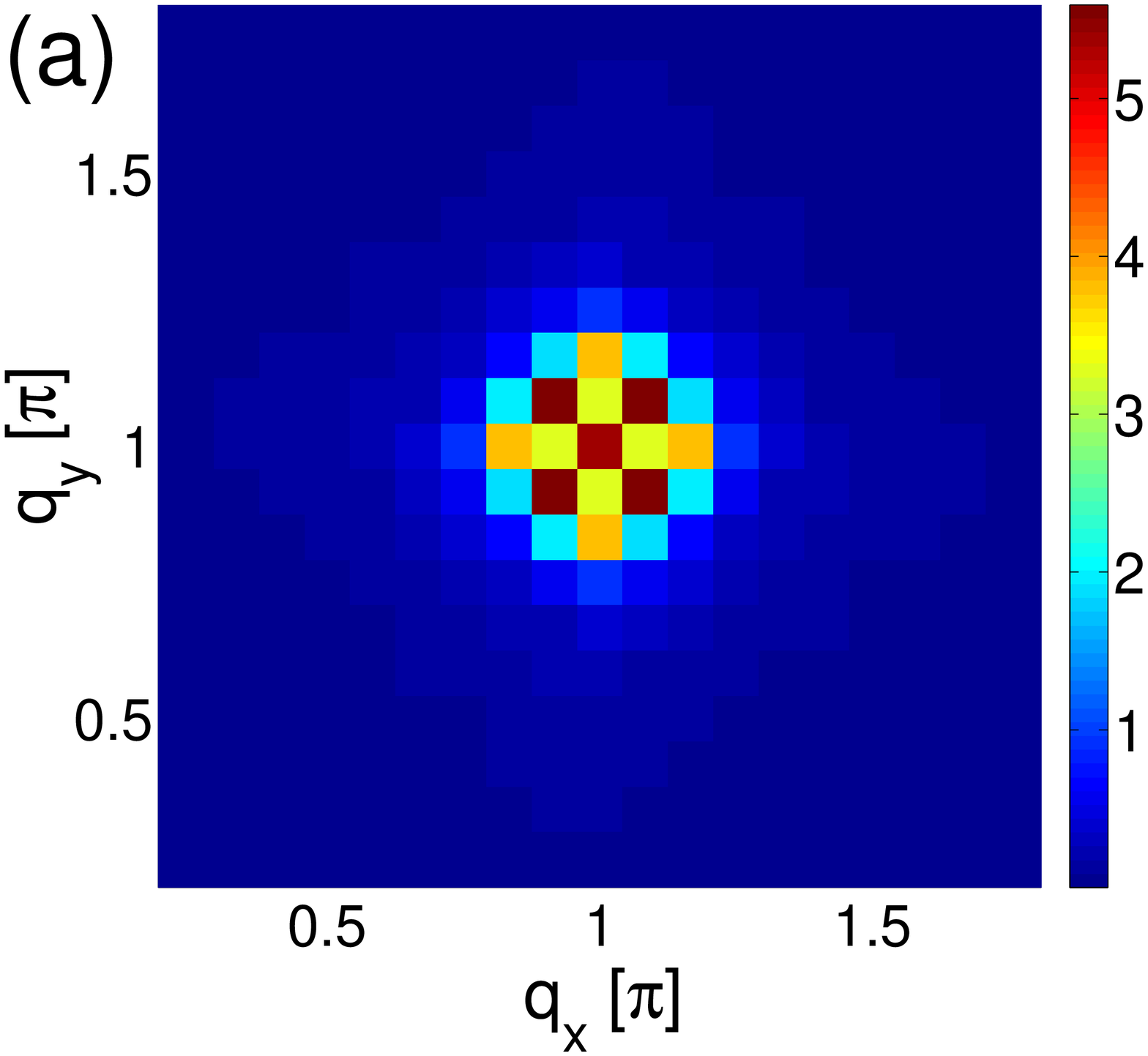}
\end{minipage}
\begin{minipage}{.49\columnwidth}
\includegraphics[clip=true,height=0.98\columnwidth,width=0.98\columnwidth]{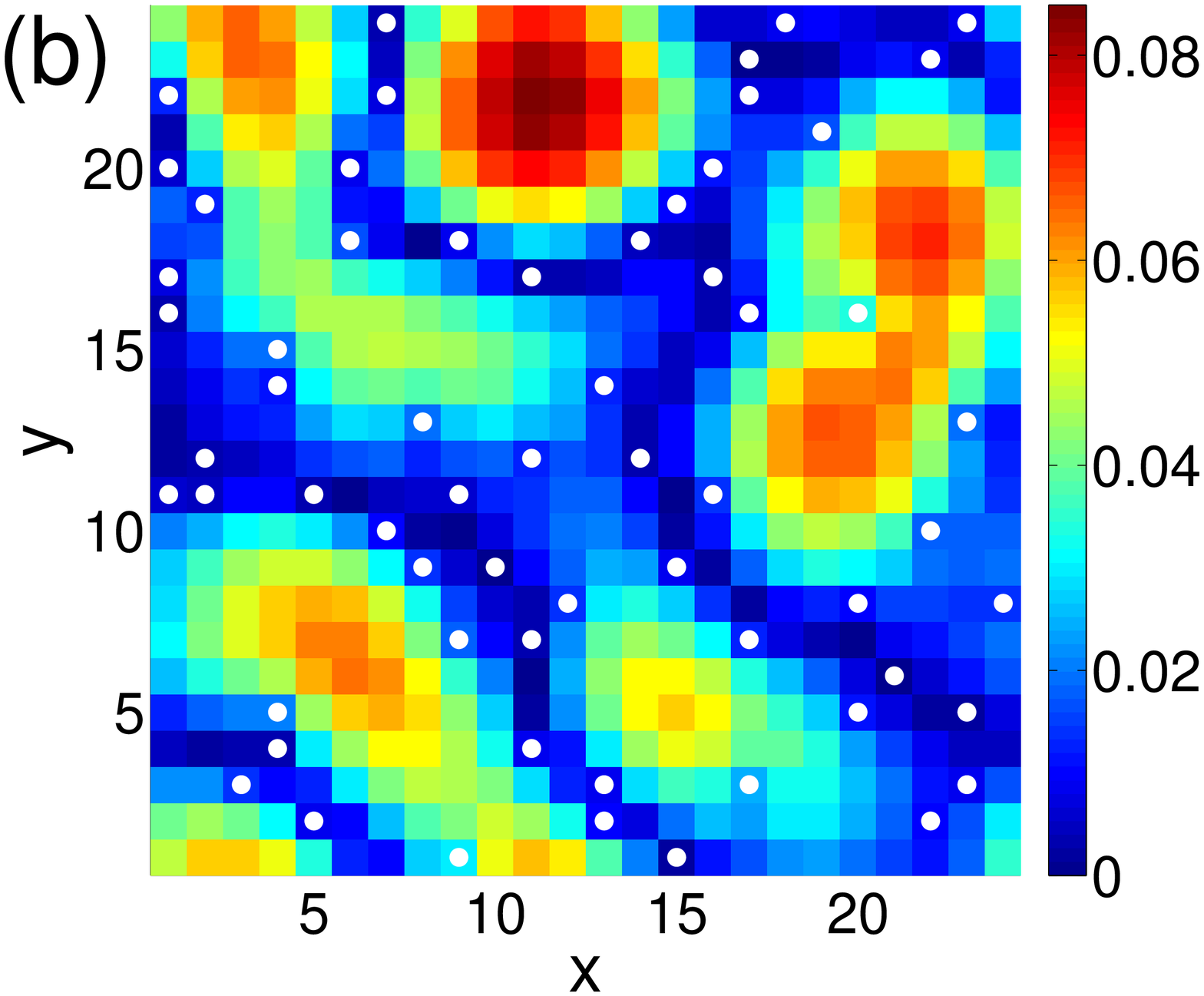}
\end{minipage}
\\
\begin{minipage}{.49\columnwidth}
\includegraphics[clip=true,height=0.98\columnwidth,width=0.98\columnwidth]{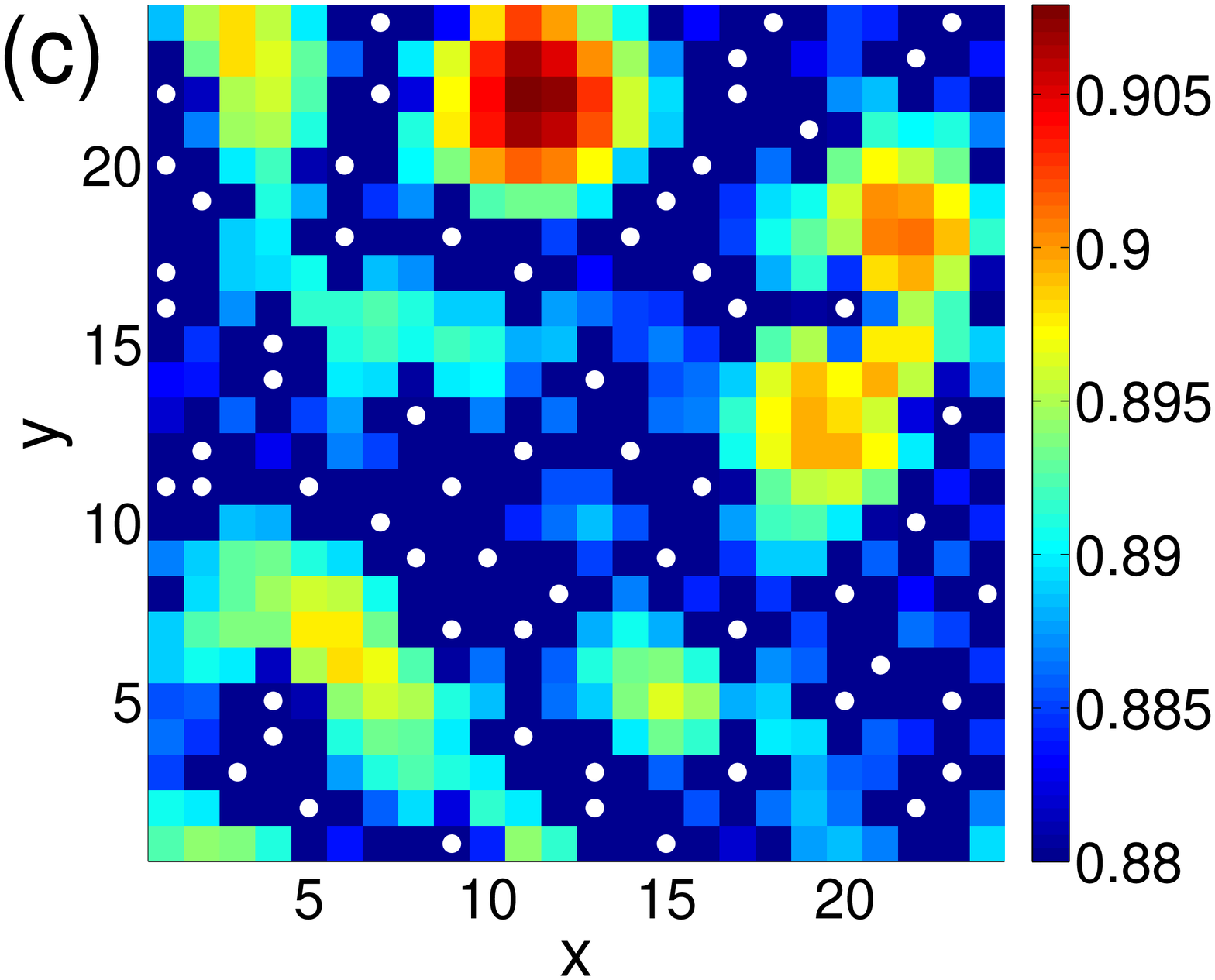}
\end{minipage}
\begin{minipage}{.49\columnwidth}
\includegraphics[clip=true,height=0.98\columnwidth,width=0.98\columnwidth]{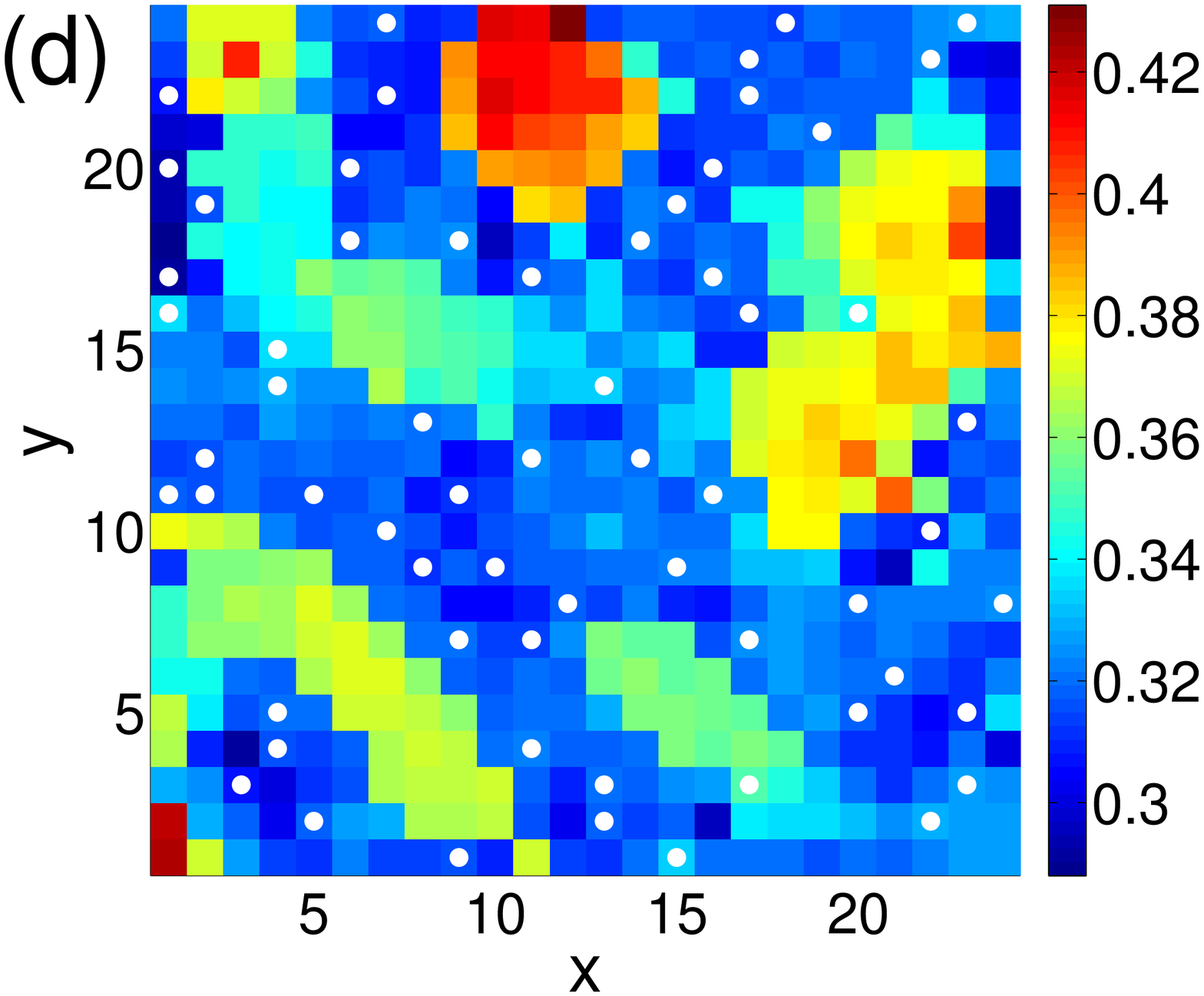}
\end{minipage}
\\
\begin{minipage}{.49\columnwidth}
\includegraphics[clip=true,height=0.98\columnwidth,width=0.98\columnwidth]{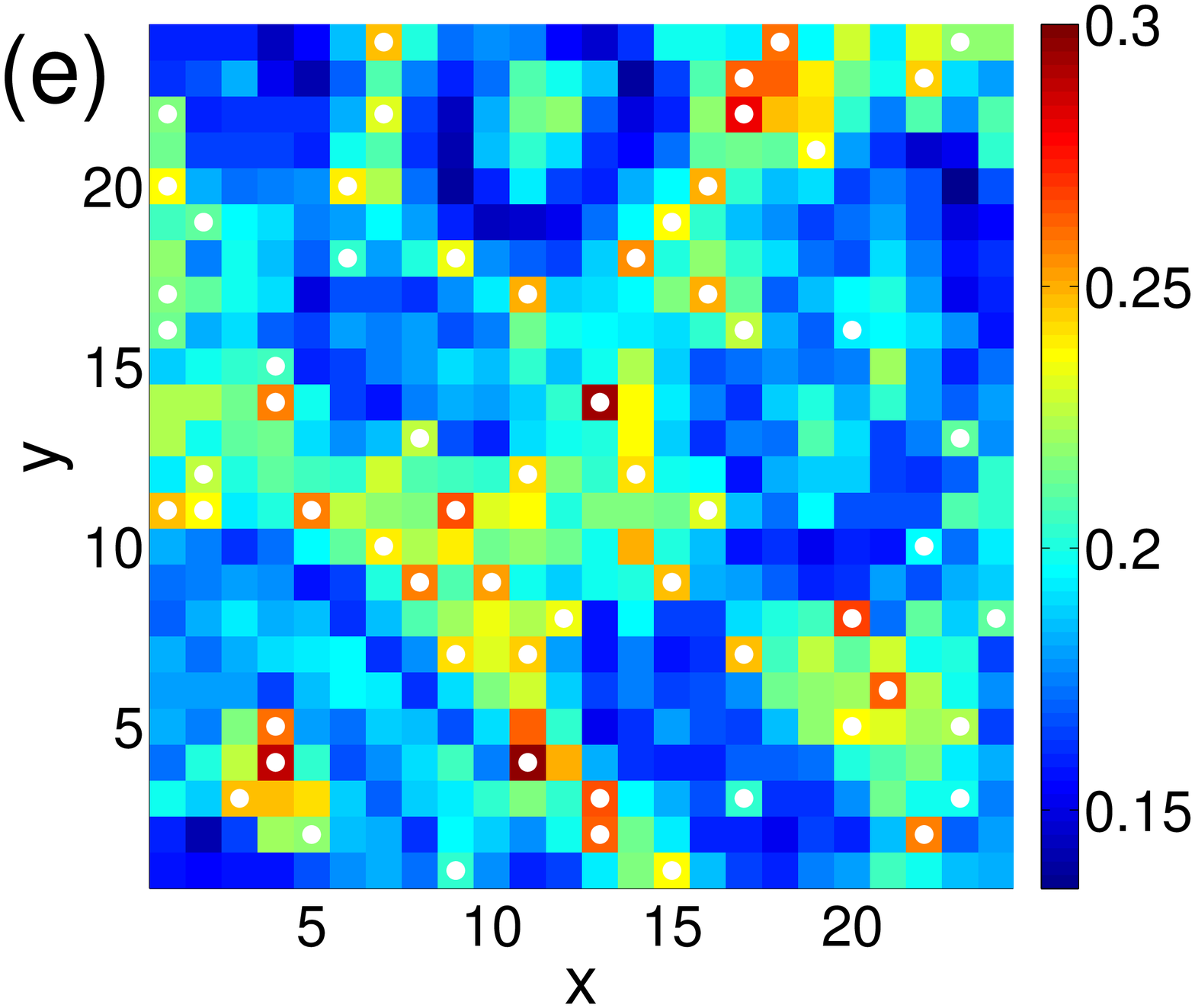}
\end{minipage}
\begin{minipage}{.49\columnwidth}
\includegraphics[clip=true,height=0.98\columnwidth,width=0.98\columnwidth]{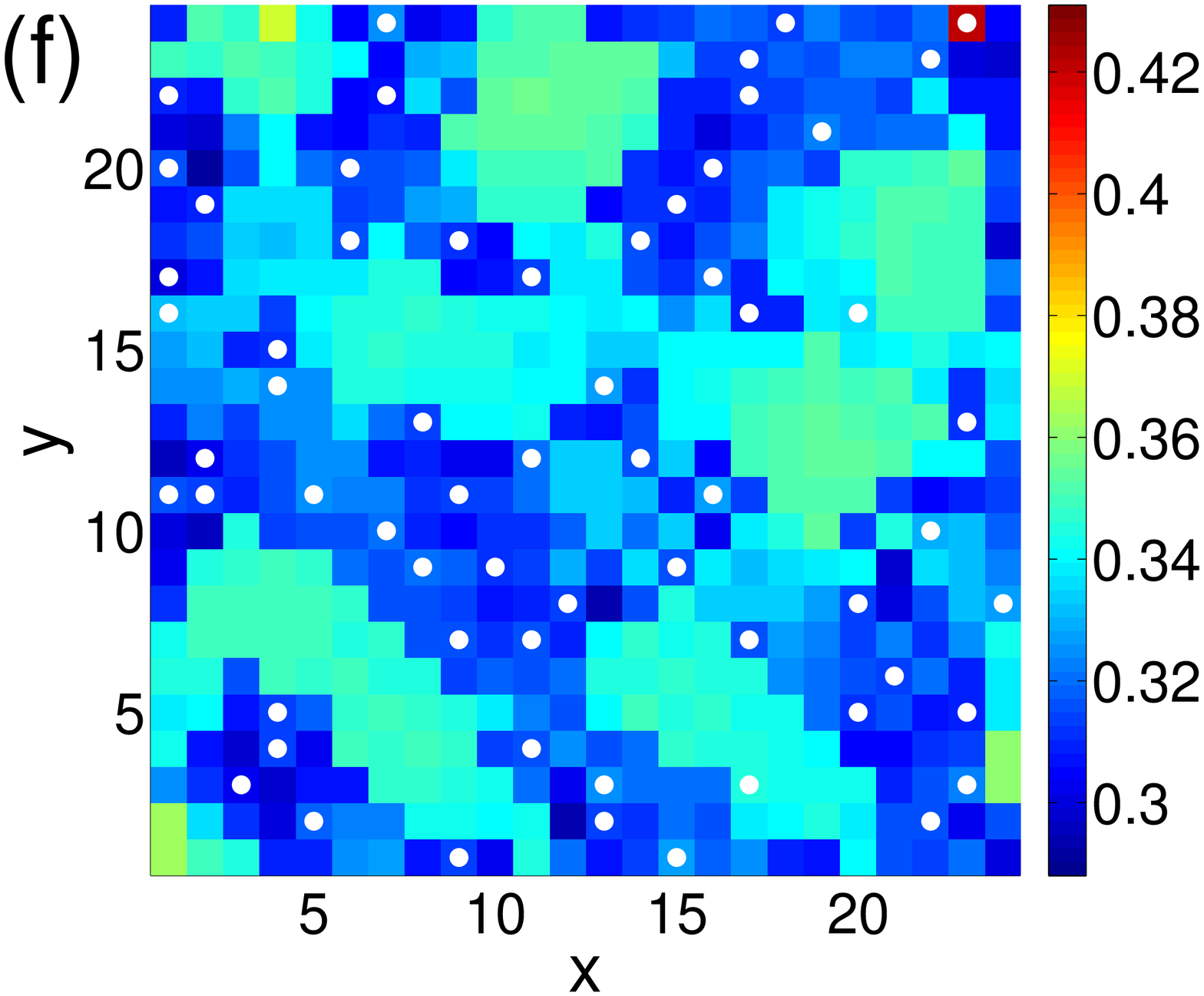}
\end{minipage}
\caption{(Color online) (a) Fourier transform of the magnetization averaged over 10 distinct impurity configurations. (b) Absolute value of the magnetization $|m_i|$. The white dots show the positions of the impurities. (c) Electronic charge density $n_i$. (d) Gapmap extracted from the LDOS (half the distance to the positive coherence peak). (e) Peak height extracted from the LDOS (positive coherence peak). (f) Gapmap extracted from the LDOS, but where the possibility for magnetic order is removed by hand. The parameters are identical to those used in Fig. \ref{figmagn}(b).} \label{figx0125}
\end{figure}

\begin{figure}[]
\begin{minipage}{.49\columnwidth}
\includegraphics[clip=true,height=0.98\columnwidth,width=0.98\columnwidth]{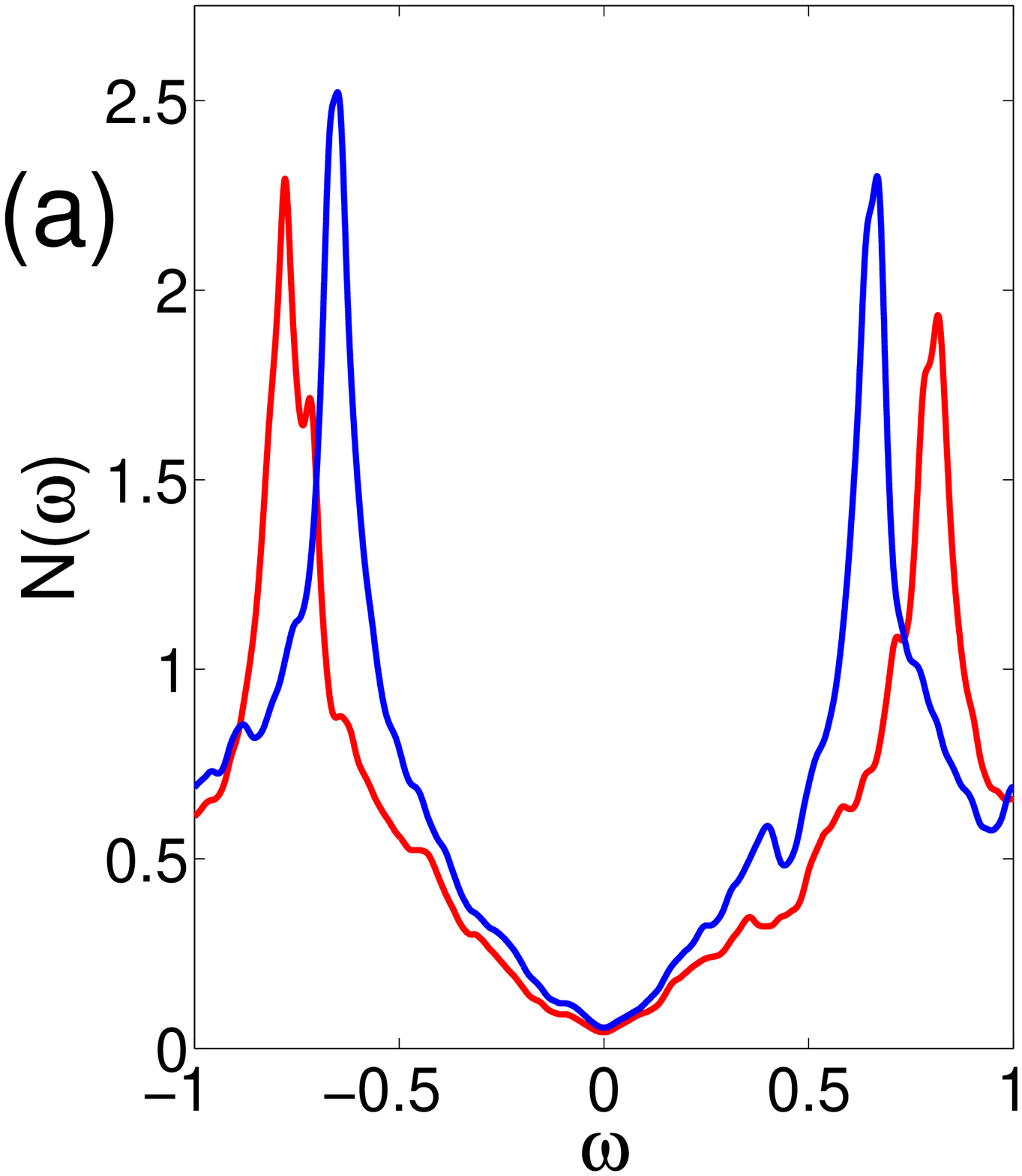}
\end{minipage}
\begin{minipage}{.49\columnwidth}
\includegraphics[clip=true,height=0.98\columnwidth,width=0.98\columnwidth]{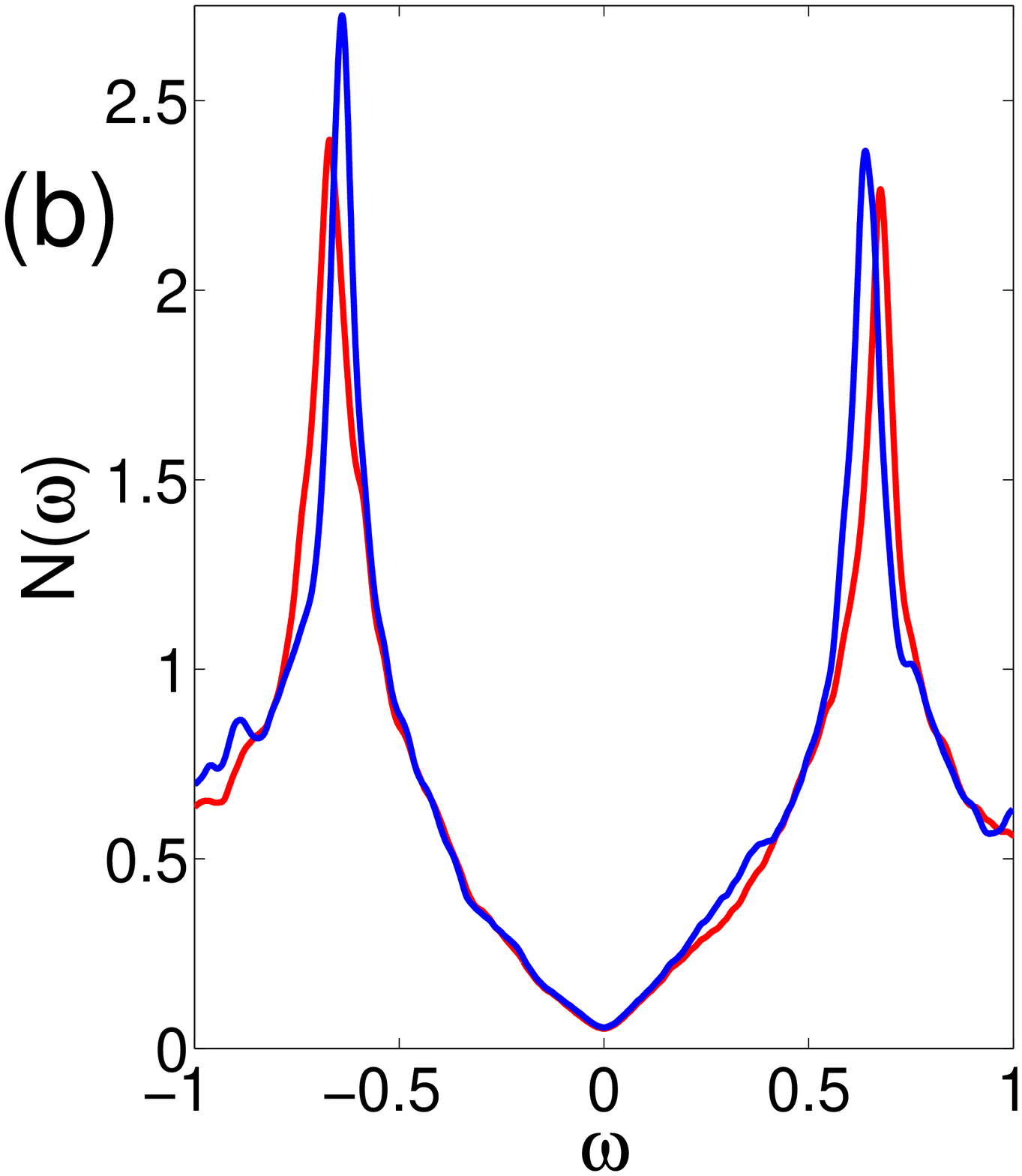}
\end{minipage}
\\
\begin{minipage}{.49\columnwidth}
\includegraphics[clip=true,height=0.98\columnwidth,width=0.98\columnwidth]{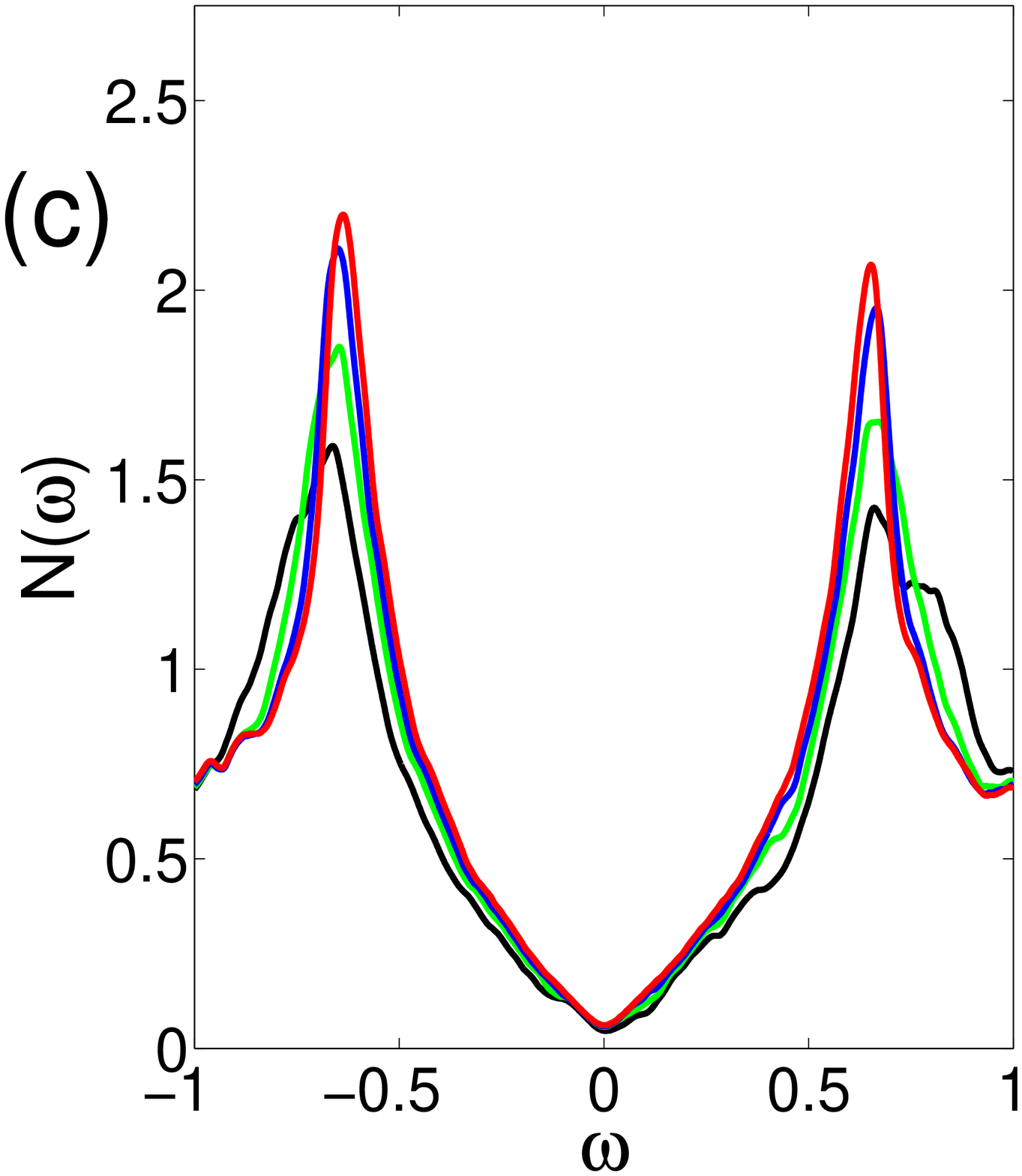}
\end{minipage}
\begin{minipage}{.49\columnwidth}
\includegraphics[clip=true,height=0.98\columnwidth,width=0.98\columnwidth]{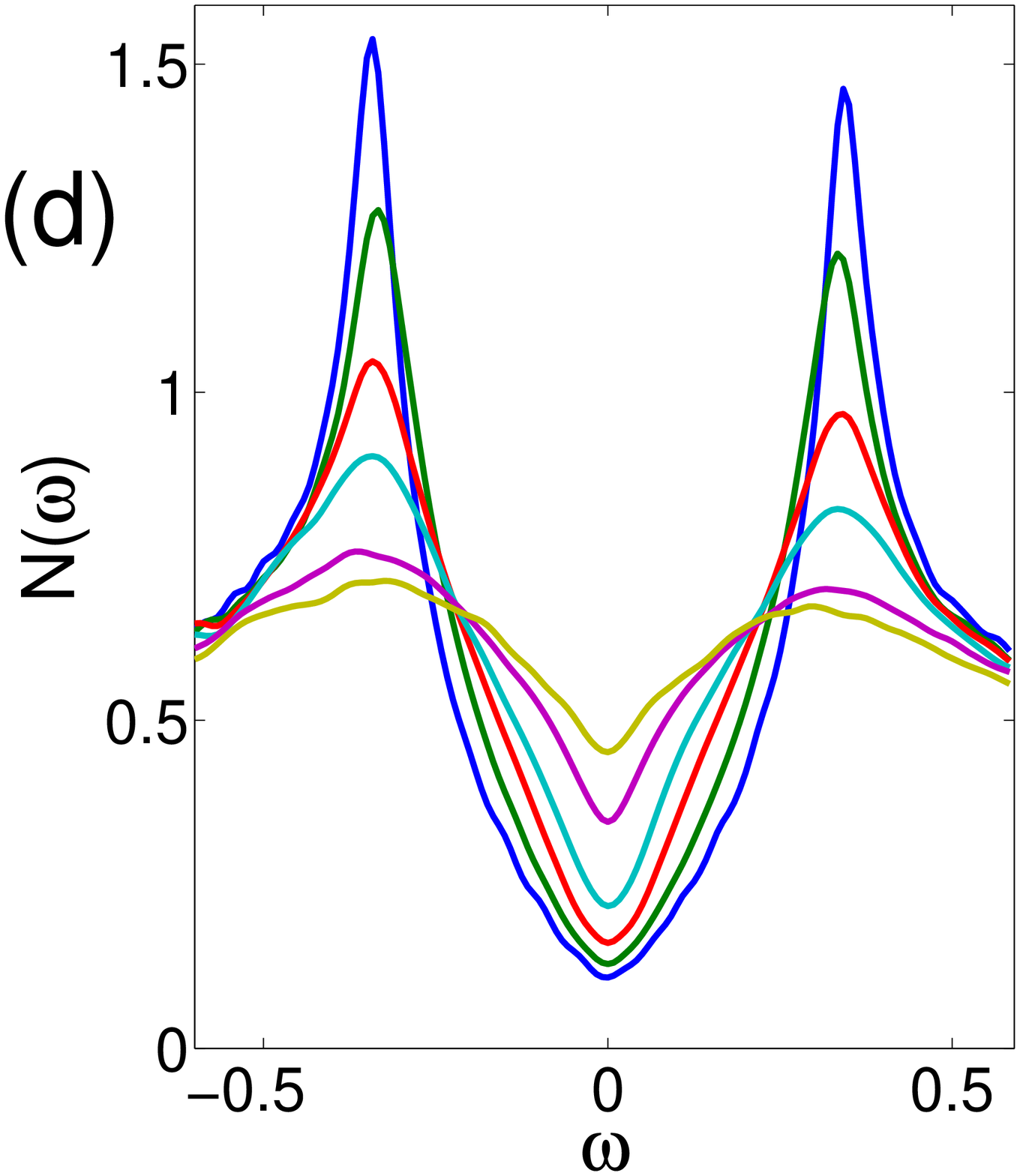}
\end{minipage}
\caption{(Color online) LDOS $N(\omega)$ vs. energy $\omega$. (a) LDOS for two sites in Fig. \ref{figmagn}(b) (red curve is from a large-gap region [site (13,21)], blue in a small-gap region [site (13,11)]. (b) LDOS for the same two sites but in Fig. \ref{figmagn}(d). (c) Spatially averaged LDOS corresponding to the four panels in Fig. \ref{figmagn} (black $\delta=0.115$, green $\delta=0.125$, blue $\delta=0.13$, and red $\delta=0.135$). (d) The spatially averaged LDOS for a dirty $d$-wave superconductor with $t'= -0.25t$, $J=1.1t$, and $n_{imp}= 0\%, 1\%, 5\%, 10\%, 20\%$ (bottom to top at $\omega=0$). The calculations were done using $20 \times 20$ or $24 \times 24$ lattices with $10\times 10$ supercells and (d) was averaged over 10 different impurity configurations. An artificial broadening of the delta-functions in Eq. \eqref{Nomega} with $\eta=0.016$ was used to smoothen the curves slightly.} \label{ldoscomp}
\end{figure}

Representative curves for $N(\omega)$ in different local gap regions are shown in Fig. \ref{ldoscomp}(a,b) where one clearly sees the anti-correlation between local gap and coherence peak height. As discussed previously, the presence of a robust 'V' shaped density of states at low energies is novel and not contained within models that ignore electronic correlations.\cite{garg,BMAndersen:2006,andersen08,tklee2009} This can be seen explicitly from Fig. \ref{ldoscomp}(c,d) where we compare the spatially averaged LDOS within the present model and a conventional Bogoliubov-de Gennes approach.\cite{garg} The latter model clearly piles up states inside the gap. By contrast, the spatially averaged LDOS shown in Fig. \ref{ldoscomp}(b) corresponding to the four panels in Fig. \ref{figmagn} displays a remarkably universal low-energy density of states. The magnetization at low doping causes additional minor structure inside the gap. We interpret these sub-gap kinks as a disordered version of a similar in-gap kink caused by altered contours of constant energies in a model with bulk coexistence of AF and $d$-wave superconducting order.\cite{Andersen08,granath} Such features could serve as tunneling fingerprints of local magnetism in the underdoped regime.

\begin{figure}[]
\begin{minipage}{.49\columnwidth}
\includegraphics[clip=true,height=0.98\columnwidth,width=0.98\columnwidth]{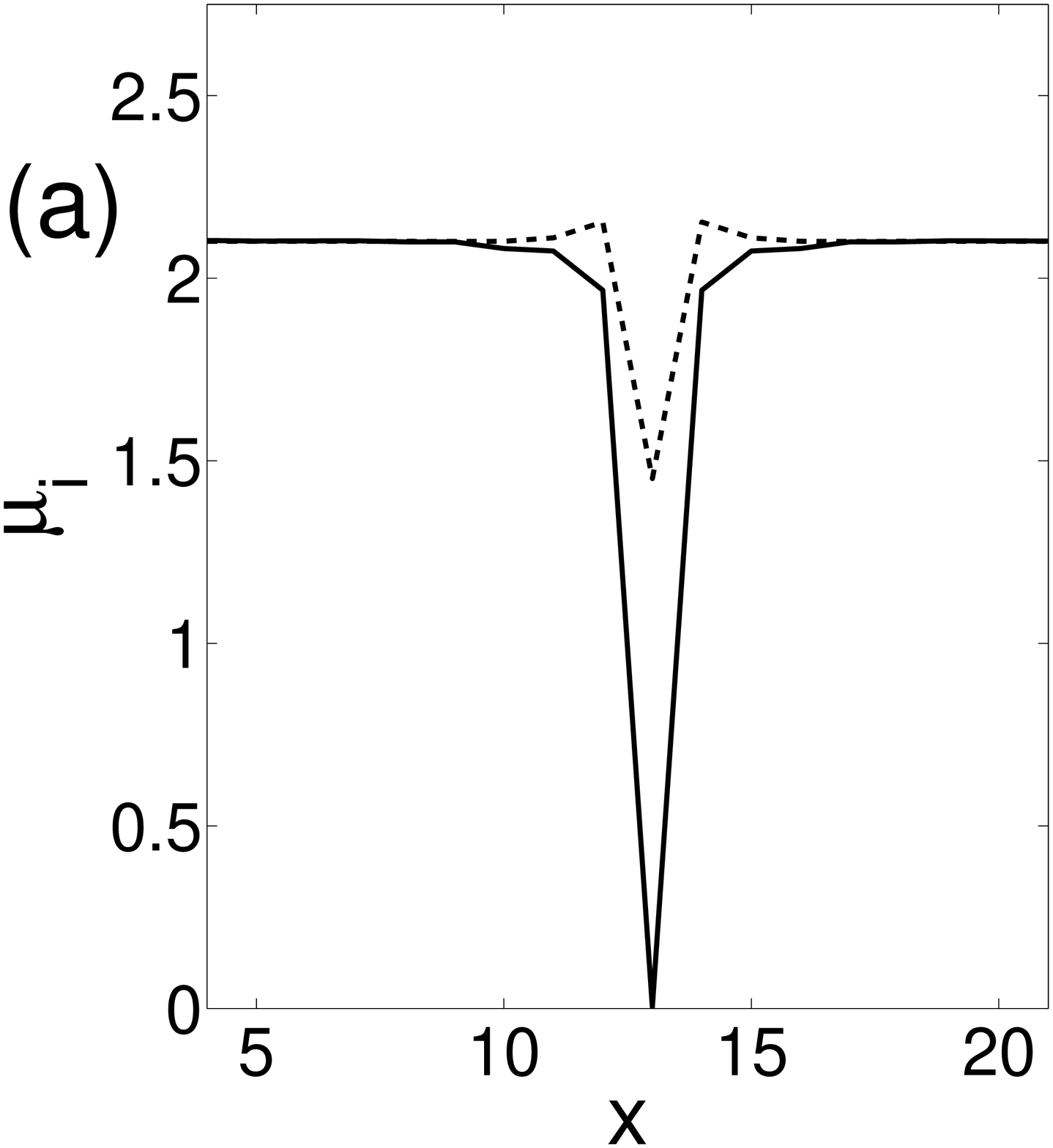}
\end{minipage}
\begin{minipage}{.49\columnwidth}
\includegraphics[clip=true,height=0.98\columnwidth,width=0.98\columnwidth]{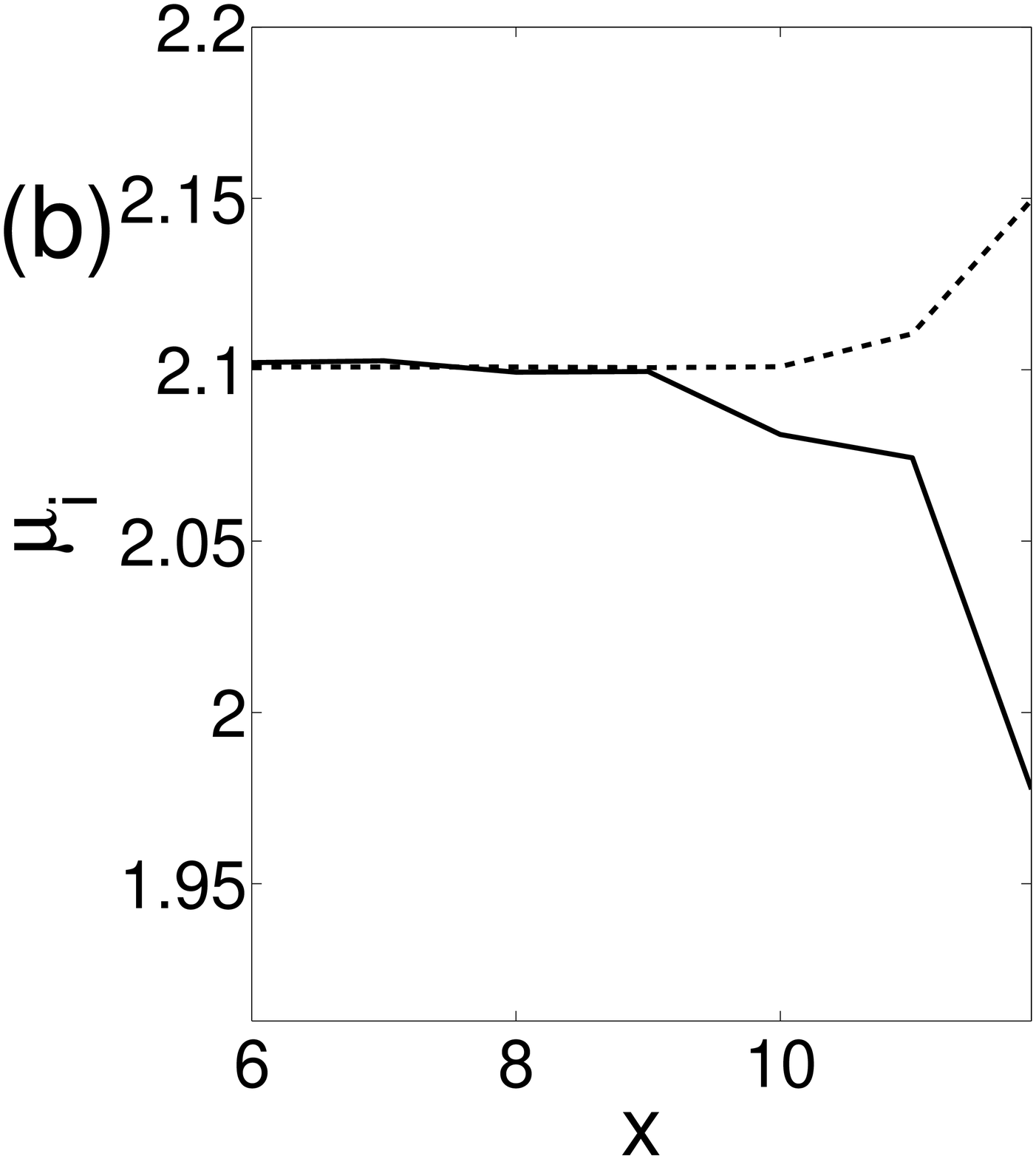}
\end{minipage}
\caption{(a) the local chemical potential $\mu_{i}$ in the $x$ direction for a $24 \times 24$ system containing a single impurity with $V_i=t$ (dashed) or $V_i=100t$ (solid). The impurity is situated at site=13 and $\delta=12.5$. (b) the same as (a) but cut off right before the impurity site to highlight the spatial dependence away from the impurity site.}\label{fig:mu}
\end{figure}

Next, to underline the importance of minimizing the energy and to understand the difference between a strong and weak scatterer we define a "local chemical potential" $\mu_{i}$ defined from
\begin{eqnarray}
\frac{\partial W}{\partial n_{i\sigma}}&=&-\left(\mu -V_{i}\right) - \mu_i.
\end{eqnarray}
Figure \ref{fig:mu} shows the local chemical potential $\mu_{i}$ for different impurity potentials. It is clear from Fig. \ref{fig:mu}(a) that $\mu_{i}$ works against the impurity potential on the impurity site since a larger $V_i$ leads to a smaller $\mu_{i}$. However, the renormalization of the  impurity potential by $\mu_{i}$ has the largest impact on weak impurity potentials because $\mu_{i}$ takes values in the range $\left[0,2.15\right]$. Figure \ref{fig:mu}(b) illustrates how $\mu_{i}$ varies for the neighboring  sites depending on the impurity potential. For weak impurities $\mu_{i}$ spreads out the impurity potential, while for large impurity potentials $\mu_{i}$ attracts electrons to the neighboring sites.
The dominant contribution to $\mu_{i}$ turns out to be $\frac{dg_{i}^{t}}{dn_{i\sigma}}$ which is related to the kinetic energy. Thus for weak impurity potentials, the holes on the impurity site tend to gain kinetic energy by spreading out as a consequence of the change in the effective hopping integral, while the opposite is the case for a strong impurity. To summarize, the strong correlations treated within the GA, affect the impurity potential in two distinct ways. First, the suppression of the electron density at the impurity site $i$ enhances the effective hopping integral $t_{eff}=g^{t}_{ij}t$ which acts as a healing of the damage done by the impurity. Second, the electron density redistributes to minimize the energy because the Gutzwiller factors themselves depend on the electron density. The redistribution of the electron density is enforced by the local chemical potential which also tends to work against the impurity potential.

\section{Conclusions}

In this work, we have studied disorder-induced magnetization within the $t-J$ model with correlations treated using the Gutzwiller approximation to implement the no-double-occupancy constraint.
 In general, the inclusion of correlations strong enough to describe band narrowing and
 other crude features of the Mott transition locally was seen to justify the phenomenological
 description of underdoping in terms of a renormalized Hubbard $U/t$ used until now in Hartree-Fock treatments of disordered correlated
 $d$-wave superconductors.  However, some unexpected subtleties were also discovered. In the case of dopant disorder where the impurity concentration equals the doping level and the individual disorder potentials are weak, local charge reorganization can induce regions with finite magnetization located away from the impurities. On the other hand, strong scatterers locally pin AF regions which may merge and eventually form a quasi-long-range ordered spatial structure. Remarkably, the LDOS at low energies remains largely disorder independent whereas the superconducting gap (coherence peak height) extracted from the LDOS is increased (decreased) in the magnetic regions present in the underdoped regime.

The goal of this analysis  is to work towards  a theory  incorporating disorder, superconductivity and correlations capable of describing local spectroscopies of cuprates across the phase diagram.  Perhaps the most successful approach thus far, in terms of reproducing the many statistical observables
 reported by STM, has been the phenomenology of Nunner {\it et al.}\cite{nunner1} To some extent this success
was only possible, however, because a) correlations were neglected and doping dependence ignored; and b) an
impurity was assumed {\it ad hoc} to add independent Coulomb and pairing potentials to the system.
The present work has attempted primarily to address the deficiency represented by a).  Earlier works, notably Ref. \onlinecite{Wang02} also incorporated disorder and correlation in the superconducting state, but neglected both the modulation of the pairing interaction by disorder and the possibility of local magnetism.
The latter effect has been shown here to lead to some of the important correlations present in $\mu$SR, neutron scattering, and STM; in particular the
anti-correlation between peak height and gap amplitude
has been exhibited in the locally ordered magnetic phases.

\begin{figure}[]
\includegraphics[clip=true,width=0.9\columnwidth]{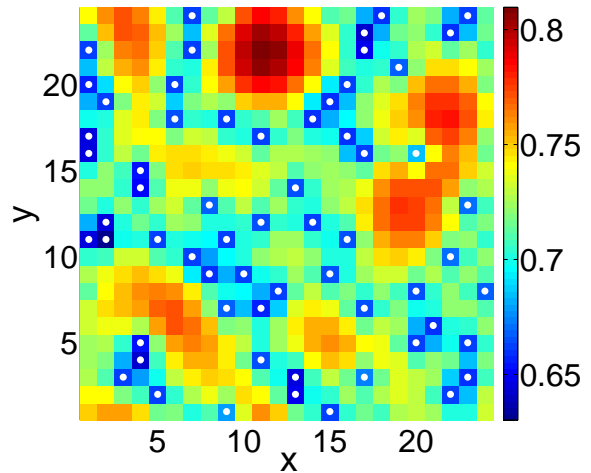}
\caption{Real-space plot of the effective exchange coupling $\tilde{J}_i=\sum_j \tilde{J}_{ij}/4$, where $j$ are the four nearest neighbors to site $i$, obtained from $\tilde{J}_{ij}=\frac{1}{2}\sum_\sigma \frac{\partial W}{\partial \Delta_\sigma}/\Delta_\sigma$ shown for a configuration of weak impurities [type I disorder] with $n_{imp}=\delta=0.125$ similar to Fig. \ref{figx0125}}\label{fig:J}
\end{figure}

On the other hand, the modulation of the pairing potential represented by the exchange
constant $J$ in the model renormalized by the local Gutzwiller factors is relatively weak, as is shown
in Fig. \ref{fig:J}.
Thus the positive correlations of the gap size with the O defect position, identified as crucial in
McElroy {\it et al.},\cite{mcelroy2005} will not occur in the present model when the correlations become weak enough in the overdoped phase. From Fig. \ref{fig:J} one can see that the impurities cause local reductions of $J$ because the Gutzwiller factor $g^s$ is a decreasing function of the doping level. The present approach, therefore, does not include the possibility, discussed in Nunner {\it et al.},\cite{nunner1} that the defect distorts the lattice locally leading to a different pairing interaction, as represented, e.g. by a local enhancement of $J$. Local impurity-enhancements of $J$ is also obtained by explicitly including the different impurity potentials on the two sites involved in the exchange process.\cite{Maska} The notion of local $J$-enhancement has been explored
 by Maska {\it et al.},\cite{Maska} Foyevstova {\it et al.},\cite{foyevtsova} Johnston {\it et al.},\cite{johnston} and Khaliullin {\it et al.}\cite{Khaliullin10} with model-dependent results; all however indicate that sizeable modulations of the pairing interaction can indeed occur. These effects have been left
 out of the current approach, but will be included in our future studies of this  problem.
 
 For completeness, we mention a second possibility\cite{Wang07} to explain the defect-gap correlations at optimal to overdoping,
that a second source of disorder is present, correlated with the O dopants, and not imaged in the
experiment of McElroy {\it et al.}\cite{mcelroy2005}  This scenario is plausible but requires the existence of a second, independent source of disorder.  In addition, it has been found in density functional
theory structural studies that both in the O dopant case and that of the structural supermodulation, CuO$_4$ half-octahedra  are tilted in
identical ways by the perturbation,\cite{He2006,He2008} and are correlated empirically the same way with the gap modulations,\cite{mcelroy2005,Slezak2008,andersensupermod} lending credence to the idea that a single set of O dopants is the primary driver of the structural distortions and gap changes.  We therefore believe that the combination of effect of electronic correlations as described here,
together with a practical description of the modulation of the pairing gap, should provide a complete description of the 
statistics of local STM observables over the whole phase diagram.

\section{Acknowledgements}

We acknowledge useful discussions with W. A. Atkinson and S. Graser.
B.M.A. acknowledges support from The Danish Council for Independent Research $|$ Natural Sciences. P.J.H acknowledges support from NSF-DMR-1005625.


\begin{thebibliography}{00}

\bibitem{cren} T. Cren, D. Roditchev, W. Sacks, and J. Klein, Europhys. Lett. {\bf 54}, 84 (2001).
%
\bibitem{howald} C. Howald, P. Fournier, and A. Kapitulnik, Phys. Rev. B {\bf 64}, 100504(R) (2001).
%
\bibitem{lang} K. M. Lang, V. Madhavan, J. E. Hoffman, E. W. Hudson, H. Eisaki, S. Uchida, and J. C. Davis, Nature (London) {\bf 415}, 412 (2002).
%
\bibitem{mcelroy} K. McElroy, D.-H. Lee, J. E. Hoffman, K. M. Lang, J. Lee, E. W. Hudson, H. Eisaki, S. Uchida, and J. C. Davis, Phys. Rev. Lett. {\bf 94}, 197005 (2005).
%
\bibitem{fang} A. C. Fang, L. Capriotti, D. J. Scalapino, S. A. Kivelson, N. Kaneko, M. Greven, and A. Kapitulnik, Phys. Rev. Lett. {\bf 96}, 017007 (2006).
%
\bibitem{gomes} K. K. Gomes, A. N. Pasupathy, A. Pushp, S. Ono, Y. Ando, and A. Yazdani, Nature (London) {\bf 447}, 569 (2007).
%
\bibitem{stock06} C. Stock, W. J. L. Buyers, Z. Yamani, C. L. Broholm, J.-H. Chung, Z. Tun, R. Liang, D. Bonn, W. N. Hardy, and R. J. Birgeneau, Phys. Rev. B {\bf 73}, 100504(R) (2006).
%
\bibitem{sonier} J. Sonier, F. D. Callaghan, Y. Ando, R. F. Kiefl, J. H. Brewer, C. V. Kaiser, V. Pacradouni, S. A. Sabok-Sayr, X. F. Sun, S. Komiya, W. N. Hardy, D. A. Bonn, and R. Liang, Phys. Rev. B {\bf 76}, 064522 (2007).
%
\bibitem{stock08} C. Stock, W. J. L. Buyers, Z. Yamani, Z. Tun, R. J. Birgeneau, R. Liang, D. Bonn, and W. N. Hardy, Phys. Rev. B {\bf 77}, 104513 (2008).
%
\bibitem{hinkov} V. Hinkov, D. Haug, B. Fauqu\'{e}, P. Bourges, Y. Sidis, A. Ivanov, C. Bernhard, C. T. Lin, and B. Keimer, Science {\bf 319}, 597 (2008).
%
\bibitem{keimer}B. Keimer, N. Belk, R. J. Birgeneau, A. Cassanho, C. Y. Chen, M. Greven, M. A. Kastner, A. Aharony, Y. Endoh, R. W. Erwin, and G. Shirane, Phys. Rev. B {\bf 46}, 14034 (1992).
%
\bibitem{wakimoto} S. Wakimoto, R. J. Birgeneau, Y. S. Lee, and G. Shirane, Phys. Rev. B {\bf 63}, 172501 (2001).
%
\bibitem{bella} B. Lake, H. M. R\o nnow, N. B. Christensen, G. Aeppli, K. Lefmann, D. F. McMorrow, P. Vorderwisch, P. Smeibidl, N. Mangkorntong, T. Sasagawa, M. Nohara, H. Takagi, and T. E. Mason, Nature (London) {\bf 415}, 299 (2002).
%
\bibitem{julien} M.-H. Julien, Physica B {\bf 329-333}, 693 (2003).
%
\bibitem{Andersen08} B. M. Andersen and P. J. Hirschfeld, Phys. Rev. B {\bf 79}, 144515 (2009).
%
\bibitem{pan} S. H. Pan, E. W. Hudson, K. M. Lang, H. Eisaki, S. Uchida, and J. C. Davis, Nature (London) {\bf 403}, 746 (2000).
%
\bibitem{hirota} K. Hirota, K. Yamada, I. Tanaka, and H. Kojima, Physica B (Amsterdam) {\bf 241-243}, 817 (1998).
%
\bibitem{kimura03} H. Kimura, M. Kofu, Y. Matsumoto, and K. Hirota, Phys. Rev. Lett. {\bf 91}, 067002 (2003).
%
\bibitem{savici} A. T. Savici, A. Fukaya, I. M. Gat-Malureanu, T. Ito, P. L. Russo, Y. J. Uemura, C. R. Wiebe, P. P. Kyriakou, G. J. MacDougall, M. T. Rovers, G. M. Luke, K. M. Kojima, M. Goto, S. Uchida, R. Kadono, K. Yamada, S. Tajima, T. Masui, H. Eisaki, N. Kaneko, M. Greven, and G. D. Gu, Phys. Rev. Lett. {\bf 95}, 157001 (2005).
%
\bibitem{sidis96} Y. Sidis, P. Bourges, B. Hennion, L. P. Regnault, R. Villeneuve, G. Collin, and J. F. Marucco, Phys. Rev. B {\bf 53}, 6811 (1996).
%
\bibitem{suchaneck} A. Suchaneck, V. Hinkov, D. Haug, L. Schulz, C. Bernhard, A. Ivanov, K. Hradil, C. T. Lin, P. Bourges, B. Keimer, and Y. Sidis, Phys. Rev. Lett. {\bf 105}, 037207 (2010).
%
\bibitem{kimura99} H. Kimura, K. Hirota, H. Matsushita, K. Yamada, Y. Endoh, S.-H. Lee, C. F. Majkrzak, R. Erwin, G. Shirane, M. Greven, Y. S. Lee, M. A. Kastner, and R. J. Birgeneau, Phys. Rev. B {\bf 59}, 6517 (1999).
%
\bibitem{mendels} P. Mendels, H. Alloul, J. H. Brewer, G. D. Morris, T. L. Duty, S. Johnston, E. J. Ansaldo, G. Collin, J. F. Marucco, C. Niedermayer, D. R. Noakes, and C. E. Stronach, Phys. Rev. B {\bf 49}, 10035 (1994).
%
\bibitem{bernhard} C. Bernhard, Ch. Niedermayer, T. Blasius, G. V. M. Williams, R. De Renzi, C. Bucci, and J. L. Tallon, Phys. Rev. B {\bf 58},  R8937 (1998).
%
\bibitem{niedermayer} Ch. Niedermayer, C. Bernhard, T. Blasius, A. Golnik, A. Moodenbaugh, and J. I. Budnick, Phys. Rev. Lett. {\bf 80}, 3843 (1998).
%
\bibitem{watanabe} I. Watanabe, T. Adachi, K. Takahashi, S. Yairi, Y. Koike, and K. Nagamine, Phys. Rev. B {\bf 65}, 180516(R) (2002).
%
\bibitem{panagopoulos} C. Panagopoulos, J. L. Tallon, B. D. Rainford, T. Xiang, J. R. Cooper, and C. A. Scott, Phys. Rev. B {\bf 66}, 064501 (2002).
%
\bibitem{kaul} M. Vojta, T. Vojta, and R. K. Kaul, Phys. Rev. Lett.  {\bf 97}, 097001 (2006).
%
\bibitem{robertson} J. A. Robertson, S. A. Kivelson, E. Fradkin, A. C. Fang, and A. Kapitulnik, Phys. Rev. B {\bf 74}, 134507 (2006).
%
\bibitem{delmaestro} A. Del Maestro, B. Rosenow, and S. Sachdev, Phys. Rev. B {\bf 74}, 024520 (2006).
%
\bibitem{BMAndersen}  B. M. Andersen, P. J. Hirschfeld, A. P. Kampf, and M. Schmid, Phys. Rev. Lett. {\bf 99}, 147002 (2007).
%
\bibitem{BMAndersen:2006} B. M. Andersen and P. J. Hirschfeld, Physica C (Amsterdam) {\bf 460-462}, 744
(2007).
%
\bibitem{alvarez} G. Alvarez, M. Mayr, A. Moreo, and E. Dagotto, Phys. Rev. B {\bf 71}, 014514 (2005).
%
\bibitem{atkinson} W. A. Atkinson, Phys. Rev. B {\bf 75}, 024510 (2007).
%
\bibitem{AndersenJPCS} B. M. Andersen, S. Graser, M. Schmid, A. P. Kampf, and P. J. Hirschfeld, J. Phys. Chem. Solids {\bf 72}, 358 (2011).
%
\bibitem{Schmid10} M. Schmid, B. M. Andersen, A. P. Kampf, and P. J. Hirschfeld, New J. Phys. {\bf 12}, 053043 (2010).
%
\bibitem{tsuchiura} H. Tsuchiura, Y. Tanaka, M. Ogata, and S. Kashiwaya, Phys. Rev. B {\bf 64}, 140501(R) (2001).
%
\bibitem{wanglee} Z. Wang and P. A. Lee, Phys. Rev. Lett. {\bf 89}, 217002 (2002).
%
\bibitem{zhu} J.-X. Zhu, I. Martin, and A. R. Bishop, Phys. Rev. Lett. {\bf 89}, 067003 (2002).
%
\bibitem{chen04} Y. Chen and C. S. Ting, Phys. Rev. Lett. {\bf 92}, 077203 (2004).
%
\bibitem{BMAndersen06} B. M. Andersen, A. Melikyan, T. S. Nunner, and P. J. Hirschfeld, Phys. Rev. Lett. {\bf 96}, 097004 (2006).
%
\bibitem{JWHarter:2007} J. W. Harter, B. M. Andersen, J. Bobroff, M. Gabay, and P. J. Hirschfeld, Phys. Rev. B {\bf 75}, 054520 (2007).
%
\bibitem{HAlloul:2009} H. Alloul, J. Bobroff, M. Gabay, and  P. J. Hirschfeld, Rev. Mod. Phys. {\bf 81}, 45 (2009).
%
\bibitem{andersen08} B. M. Andersen and P. J. Hirschfeld, Phys. Rev. Lett. {\bf 100}, 257003 (2008).
%
\bibitem{weichen} W. Chen, B. M. Andersen, and P. J. Hirschfeld, Phys. Rev. B {\bf 80}, 134518 (2009).
%
\bibitem{andersen10} B. M. Andersen, S. Graser, and P. J. Hirschfeld, Phys. Rev. Lett. {\bf 105}, 147002 (2010).
%
\bibitem{Himeda1} A. Himeda and M. Ogata, Phys. Rev. B \textbf{60}, R9935 (1999).
%
\bibitem{Ogata1} M. Ogata and A. J. Himeda, J. Phys. Soc. Jpn. {\bf 72}, 374 (2003).
%
\bibitem{tsuchiura2} H. Tsuchiura,  M. Ogata, Y. Tanaka, Y. Tanaka, and S. Kashiwaya, Phys. Rev. B {\bf 68}, 012509 (2003).
%
\bibitem{tsuchiura3} H. Tsuchiura and M. Ogata, Journal of Physics: Conf. Series {\bf 150}, 052272 (2009).
%
\bibitem{Wing-Ho} W.-H. Ko, C. P. Nave, and P. A. Lee, Phys. Rev. B {\bf 76}, 245113 (2007).
%
\bibitem{Yang} K.-Y. Yang, W. Q. Chen, T. M. Rice, M. Sigrist, and F.-C. Zhang, New J. Phys. {\bf 11}, 055053 (2009).
%
\bibitem{garg} A. Garg, M. Randeria, and N. Trivedi, Nature Phys. {\bf 4}, 762 (2008).
%
\bibitem{tklee2009} N. Fukushima, C.-P. Chou, and T. K. Lee, Phys. Rev. B {\bf 79}, 184510 (2009).
%
\bibitem{Zhang1} F.-C. Zhang, C. Gros, T. M. Rice, and H. Shiba, Supercond. Sci. Technol. \textbf{1}, 36 (1988).
%
\bibitem{Barash} B. M. Andersen, I. V. Bobkova, P. J. Hirschfeld, and Yu. S. Barash, Phys. Rev. B {\bf 72}, 184510 (2005).
%
\bibitem{Chen} W. Chen, K. Yang, T. M. Rice, and F. C. Zhang, Euro. Phys. Lett. {\bf 82}, 17004 (2008).
%
\bibitem{atkinson2000} W. A. Atkinson, P. J. Hirschfeld, A. H. MacDonald, and K. Ziegler, Phys. Rev. Lett. {\bf 85}, 3926 (2000).
%
\bibitem{morr} D. Morr and N.A. Stavropoulos, Phys. Rev. B {\bf 66}, 140508 (2002).
%
\bibitem{anderseninterference} B. M. Andersen and P. Hedeg\aa rd, Phys. Rev. B {\bf 67}, 172505 (2003).
%
\bibitem{mcelroy2005} K. McElroy, J. Lee, J. A. Slezak, D.-H. Lee, H. Eisaki, S. Uchida, and J. C. Davis, Science {\bf 309}, 1048 (2005).
%
\bibitem{nunner1} T. S. Nunner, B. M. Andersen, A. Melikyan, and P. J. Hirschfeld, Phys. Rev. Lett. {\bf 95}, 177003 (2005).
%
\bibitem{nunner2} T. S. Nunner, W. Chen, B. M. Andersen, A. Melikyan, and P. J. Hirschfeld, Phys. Rev. B {\bf 73}, 104511 (2006).
%
\bibitem{andersentherm} B. M. Andersen, A. Melikyan, T. S. Nunner, and P. J. Hirschfeld, Phys. Rev. B {\bf 74}, 060501(R) (2006).
%
\bibitem{granath} M. Granath and B. M. Andersen, Phys. Rev. B {\bf 81}, 024501 (2010).
%
\bibitem{Wang02}  Z. Wang, J. R. Engelbrecht, S. Wang, H. Ding, and S. H. Pan, Phys. Rev. B {\bf 65}, 064509 (2002).
%
\bibitem{Maska} M. M. Maska, Z. Sledz, K. Czajka, and M. Mierzejewski, Phys. Rev. Lett. {\bf 99}, 147006 (2007).
%
\bibitem{foyevtsova} K. Foyevtsova, R. Valent\'{\i}, and P. J. Hirschfeld, Phys. Rev. B {\bf 79}, 144424 (2009).
%
\bibitem{johnston} S. Johnston, F. Vernay, B. Moritz, Z.-X. Shen, N. Nagaosa, J. Zaanen, and T. P. Devereaux, Phys. Rev. B {\bf 82}, 064513 (2010).
%
\bibitem{Khaliullin10} G. Khaliullin, M. Mori, T. Tohyama, and S. Maekawa, Phys. Rev. Lett. {\bf 105}, 257005 (2010).
%
\bibitem{Wang07} S. Zhou, H. Ding, and Z. Wang, Phys. Rev. Lett. {\bf 98}, 076401 (2007).
%
\bibitem{He2006} Y. He, T. S. Nunner, P. J. Hirschfeld, and H.-P. Cheng, Phys. Rev. Lett. {\bf 96}, 197002 (2006).
%
\bibitem{He2008}  Y. He, S. Graser, P. J. Hirschfeld, and H.-P. Cheng, Phys. Rev. B {\bf 77}, 220507(R) (2008).
%
\bibitem{Slezak2008} J.A. Slezak, Jinho Lee, M. Wang, K. McElroy, K. Fujita, H. Eisaki, S. Uchida, B. M. Andersen, P. J. Hirschfeld, and J.C. Davis,  Proc. Natl. Acad. Sci. {\bf 105}, 3203 (2008).
%
\bibitem{andersensupermod} B. M. Andersen, P. J. Hirschfeld, and J. A. Slezak, Phys. Rev. B {\bf 76}, 020507(R) (2007).
%

\end{thebibliography}
\end{document}